\documentclass[times,twocolumn]{aastex62}

\usepackage{CJK}
\usepackage{amsmath}
\usepackage{multirow}
\usepackage{cases}
\usepackage{graphicx}
\usepackage{subfigure}
\usepackage{natbib}
\usepackage{color}
\usepackage{tabularx}
\usepackage{bm}
\usepackage{threeparttable}
\usepackage{gensymb}
\usepackage{booktabs}

\newcommand{\thetae}{\theta_{\rm E}}

\newcommand{\pie}{\pi_{\rm E}}

\newcommand{\te}{t_{\rm E}}
\newcommand{\eventa}{KMT-2025-BLG-0811}
\newcommand{\eventb}{KMT-2025-BLG-0912}

\DeclareGraphicsExtensions{.pdf,.png,.jpg}

\shorttitle{}
\shortauthors{Tang et al.}

\begin{document}
\begin{CJK*}{UTF8}{gbsn}

\title{{\large Two Low Mass-Ratio Microlensing Planets and Two Types of Central-Resonant Degeneracy}}

\correspondingauthor{Yuchen Tang, Weicheng Zang}
\email{tangyuchen@westlake.edu.cn, zangweicheng@westlake.edu.cn}

\author[0000-0001-5651-9440]{Yuchen Tang (唐雨辰)}
\affiliation{Department of Astronomy, Westlake University, Hangzhou 310030, Zhejiang Province, China}

\author[0000-0001-6000-3463]{Weicheng Zang (臧伟呈)}
\affiliation{Department of Astronomy, Westlake University, Hangzhou 310030, Zhejiang Province, China}

\author[0000-0001-9823-2907]{Yoon-Hyun Ryu} 
\affiliation{Korea Astronomy and Space Science Institute, Daejeon 34055, Republic of Korea}

\author[0000-0001-5207-5619]{Andrzej Udalski}
\affiliation{Astronomical Observatory, University of Warsaw, Al. Ujazdowskie 4, 00-478 Warszawa, Poland}

\author[0000-0003-0626-8465]{Hongjing Yang (杨弘靖)}
\affiliation{Westlake Institute for Advanced Study, Hangzhou 310030, Zhejiang Province, China}
\affiliation{Department of Astronomy, Westlake University, Hangzhou 310030, Zhejiang Province, China}

\collaboration{(Leading Authors)}

\author[0000-0003-3316-4012]{Michael D. Albrow}
\affiliation{University of Canterbury, School of Physical and Chemical Sciences, Private Bag 4800, Christchurch 8020, New Zealand}

\author[0000-0001-6285-4528]{Sun-Ju Chung}
\affiliation{Korea Astronomy and Space Science Institute, Daejeon 34055, Republic of Korea}

\author{Andrew Gould} 
\affiliation{Max-Planck-Institute for Astronomy, K\"onigstuhl 17, 69117 Heidelberg, Germany}
\affiliation{Department of Astronomy, Ohio State University, 140 W. 18th Ave., Columbus, OH 43210, USA}

\author[0000-0002-2641-9964]{Cheongho Han}
\affiliation{Department of Physics, Chungbuk National University, Cheongju 28644, Republic of Korea}

\author[0000-0002-9241-4117]{Kyu-Ha Hwang}
\affiliation{Korea Astronomy and Space Science Institute, Daejeon 34055, Republic of Korea}

\author[0000-0002-0314-6000]{Youn Kil Jung}
\affiliation{Korea Astronomy and Space Science Institute, Daejeon 34055, Republic of Korea}
\affiliation{National University of Science and Technology (UST), Daejeon 34113, Republic of Korea}

\author[0000-0002-4355-9838]{In-Gu Shin}
\affiliation{Department of Astronomy, Westlake University, Hangzhou 310030, Zhejiang Province, China}


\author[0000-0003-1525-5041]{Yossi Shvartzvald}
\affiliation{Department of Particle Physics and Astrophysics, Weizmann Institute of Science, Rehovot 7610001, Israel}

\author[0000-0001-9481-7123]{Jennifer C. Yee}
\affiliation{Center for Astrophysics $|$ Harvard \& Smithsonian, 60 Garden St., Cambridge, MA 02138, USA}

\author{Dong-Jin Kim}
\affiliation{Korea Astronomy and Space Science Institute, Daejeon 34055, Republic of Korea}

\author[0000-0003-0043-3925]{Chung-Uk Lee}
\affiliation{Korea Astronomy and Space Science Institute, Daejeon 34055, Republic of Korea}

\author[0000-0002-6982-7722]{Byeong-Gon Park}
\affiliation{Korea Astronomy and Space Science Institute, Daejeon 34055, Republic of Korea}

\collaboration{(The KMTNet Collaboration)}

\author{Leandro de Almeida}
\affiliation{Instituto Nacional de Pesquisas Espaciais (INPE), Avda. dos Astronautas, 1758, S\~ao Jos\'e dos Campos, Brazil}
\affiliation{SOAR Telescope / NSF’s NOIRLab, Avda. Juan Cisternas 1500, 1700000, La Serena, Chile}

\author{Yunyi Tang}
\affiliation{Department of Astronomy, Tsinghua University, Beijing 100084, China}

\author[0000-0002-1287-6064]{Zhixing Li (李知行)}
\affiliation{Department of Astronomy, Westlake University, Hangzhou 310030, Zhejiang Province, China}

\author[0000-0002-1279-0666]{Jiyuan Zhang (张纪元)}
\affiliation{Department of Astronomy, Tsinghua University, Beijing 100084, China}

\author[0009-0001-6584-7187]{Hongyu Li (李弘禹)}
\affiliation{Department of Astronomy, Tsinghua University, Beijing 100084, China}

\author[0000-0001-8317-2788]{Shude Mao (毛淑德)}
\affiliation{Department of Astronomy, Westlake University, Hangzhou 310030, Zhejiang Province, China}

\author[0000-0003-4625-8595]{Qiyue Qian}
\affiliation{Department of Astronomy, Westlake University, Hangzhou 310030, Zhejiang Province, China}
\affiliation{Department of Astronomy, Tsinghua University, Beijing 100084, China}

\author{Dan Maoz}
\affiliation{School of Physics and Astronomy, Tel-Aviv University, Tel-Aviv 6997801, Israel}

\author{Christian Elias Borges}
\affiliation{Federal University of Uberl\^andia, 
Av.~Jo\~ao Naves de \'Avila, 
ZIP Code 38408-100, 
Uberl\^andia, MG, Brazil}

\author{Fabr\'icio Santos Kalaki}
\affiliation{Federal University of Uberl\^andia, 
Av.~Jo\~ao Naves de \'Avila, 
ZIP Code 38408-100, 
Uberl\^andia, MG, Brazil}

\author{Altair Ramos Gomes J\'unior}
\affiliation{Federal University of Uberl\^andia, 
Av.~Jo\~ao Naves de \'Avila, 
ZIP Code 38408-100, 
Uberl\^andia, MG, Brazil}
\affiliation{Interinstitutional Laboratory for e-Astronomy (LIneA), Av. Pastor Martin Luther King Jr. 126, 20765-000, Rio de Janeiro, RJ, Brazil}

\author[0000-0003-4027-4711]{Wei Zhu (祝伟)}
\affiliation{Department of Astronomy, Tsinghua University, Beijing 100084, China}

\collaboration{(The MAP \& $\mu$FUN Follow-up Team)}

\author[0000-0001-7016-1692]{Przemek Mr\'{o}z}
\affiliation{Astronomical Observatory, University of Warsaw, Al. Ujazdowskie 4, 00-478 Warszawa, Poland}

\author[0000-0002-0548-8995]{Micha{\l}~K. Szyma\'{n}ski}
\affiliation{Astronomical Observatory, University of Warsaw, Al. Ujazdowskie 4, 00-478 Warszawa, Poland}

\author[0000-0002-2335-1730]{Jan Skowron}
\affiliation{Astronomical Observatory, University of Warsaw, Al. Ujazdowskie 4, 00-478 Warszawa, Poland}

\author[0000-0002-9245-6368]{Radoslaw Poleski}
\affiliation{Astronomical Observatory, University of Warsaw, Al. Ujazdowskie 4, 00-478 Warszawa, Poland}

\author[0000-0002-7777-0842]{Igor Soszy\'{n}ski}
\affiliation{Astronomical Observatory, University of Warsaw, Al. Ujazdowskie 4, 00-478 Warszawa, Poland}

\author[0000-0002-2339-5899]{Pawe{\l} Pietrukowicz}
\affiliation{Astronomical Observatory, University of Warsaw, Al. Ujazdowskie 4, 00-478 Warszawa, Poland}

\author[0000-0003-4084-880X]{Szymon Koz{\l}owski}
\affiliation{Astronomical Observatory, University of Warsaw, Al. Ujazdowskie 4, 00-478 Warszawa, Poland}

\author[0000-0002-9326-9329]{Krzysztof A. Rybicki}
\affiliation{Astronomical Observatory, University of Warsaw, Al. Ujazdowskie 4, 00-478 Warszawa, Poland}
\affiliation{Department of Particle Physics and Astrophysics, Weizmann Institute of Science, Rehovot 76100, Israel}

\author[0000-0002-6212-7221]{Patryk Iwanek}
\affiliation{Astronomical Observatory, University of Warsaw, Al. Ujazdowskie 4, 00-478 Warszawa, Poland}

\author[0000-0001-6364-408X]{Krzysztof Ulaczyk}
\affiliation{Department of Physics, University of Warwick, Gibbet Hill Road, Coventry, CV4~7AL,~UK}

\author[0000-0002-3051-274X]{Marcin Wrona}
\affiliation{Astronomical Observatory, University of Warsaw, Al. Ujazdowskie 4, 00-478 Warszawa, Poland}
\affiliation{Villanova University, Department of Astrophysics and Planetary Sciences, 800 Lancaster Ave., Villanova, PA 19085, USA}

\author[0000-0002-1650-1518]{Mariusz Gromadzki}
\affiliation{Astronomical Observatory, University of Warsaw, Al. Ujazdowskie 4, 00-478 Warszawa, Poland}

\author{Mateusz J. Mr\'{o}z}
\affiliation{Astronomical Observatory, University of Warsaw, Al. Ujazdowskie 4, 00-478 Warszawa, Poland}

\collaboration{(The OGLE Collaboration)}

\begin{abstract}
We present observations and analysis of two low planet/host mass-ratio ($q$) microlensing planets discovered in high-magnification events. KMT-2025-BLG-0811Lb has $q \sim 4.5 \times 10^{-5}$, and a Bayesian analysis favors a super-Earth/mini-Neptune orbiting an M- or K-dwarf host at a projected separation of $\sim 3$ au. KMT-2025-BLG-0912Lb has $q = 2.6 \times 10^{-4}$ and likely hosts a super-Earth/mini-Neptune around either a low-mass M dwarf or a brown dwarf at $\sim 1$ au. Even with an observing cadence of $\Gamma > 30~{\rm hr}^{-1}$ during the planetary signal, KMT-2025-BLG-0811 still exhibits the ``central-resonant'' degeneracy. Reviewing nine such events, we find that the ``central-resonant'' degeneracy can be divided into two distinct types that occupy separate regions in the plane of $q$ and normalized source radius ($\rho$). Type~I events have similar $q$ but substantially different $\rho$ and are more difficult to resolve from the light curves. For Type~II events, the ``resonant'' solutions have relatively lower $q$ and larger $\rho$. Our review provides guidance for searching for the alternative solution once one solution has been identified.

\end{abstract}

\section{Introduction}\label{intro}

The phenomenon of gravitational microlensing, first theorized by \citet{Einstein1936}, occurs when a foreground lens star aligns closely with a distant background source star. The gravity of the lens acts as a natural telescope, magnifying the light from the source and providing a powerful tool for detecting exoplanets \citep{Shude1991, Andy1992}. If a lens star hosts a planet, the gravitational field of the planet might further perturb the source light, which causes anomalies in the otherwise symmetric magnification light curve. These anomalies are most produced when the source trajectory passes through or near the caustics \footnote{Caustics refer to the locus of point in the source plane where the magnification of a point source is formally infinite, typically forming closed curves.}.

To date, microlensing has detected over 260 exoplanets \citep{NASAExo}, offering a demographic sample that complements other techniques. Unlike the transit method, which is most sensitive to short-period planets ($\lesssim 1$ au), microlensing excels at detecting planets near the Einstein ring of the host star, which typically corresponds to a physical distance of approximately 2--3 au for Galactic bulge events. This region often coincides with the water snow line \citep{snowline}, where planet formation theories such as core accretion predict an abundance of giant and cold rocky planets \citep{Mordasini2009,Ida2013}. Furthermore, because microlensing depends on mass rather than luminosity, it is uniquely capable of detecting planets around all types of host stars \citep{Mao2012,Gaudi2012}.

Statistical studies based on microlensing surveys have begun to constrain the planetary mass-ratio ($q$) function \citep{mufun,Cassan2012,Suzuki2016,Wise,OGLE_wide,OB160007}. While \citet{Suzuki2016} reported evidence for a break in the distribution at $q \sim 1.7 \times 10^{-4}$ based on a sample of 22 planets, the low-mass end of the distribution ($q < 10^{-4}$) has remained poorly constrained owing to the limited sample size. A more recent study based on a sample of 63 planets suggested a more complex, possibly bimodal, mass-ratio distribution, including the lowest-$q$ planet detected to date, OGLE-2016-BLG-0007Lb ($q \sim 6.9 \times 10^{-6}$; \citealt{OB160007}). This distribution shows no evidence for a break near the value proposed by \citet{Suzuki2016}. Although the \cite{OB160007} sample includes 20 planets with $q < 10^{-4}$, further expanding the sample of low-$q$ planets is essential for validating the mass-ratio distribution and for understanding the diversity of planetary systems revealed by microlensing.

However, detecting low-$q$ planets is not easy because their signals typically last for only a few hours. To reliably characterize such short-lived anomalies, high-cadence monitoring is indispensable. An efficient strategy is the follow-up of high-magnification (HM) events, wherein the source trajectory probes the central caustic region near the host star. Because planetary sensitivity in HM events is concentrated and predictable near the peak, intensive observations can be focused within a short window, maximizing efficiency while minimizing resources \citep{Griest1998,OB050071}.

Las Cumbres Observatory Global Telescope (LCOGT) network, consisting of 1~m telescopes equipped with Sinistro cameras \citep{LCOGT}, provides continuous, high-cadence follow-up observations that are temporally complementary to wide-field surveys by the Korean Microlensing Telescope Network (KMTNet; \citealt{KMT2016}), the Optical Gravitational Lensing Experiment (OGLE; \citealt{OGLEIV}), the Microlensing Observations in Astrophysics (MOA; \citealt{Sako2008}), and the Prime Focus Infrared Microlensing Experiment (PRIME; \citealt{PRIME}). Since July 2020, the Microlensing Astronomy Probe (MAP) collaboration has utilized the LCOGT network to conduct systematic follow-up observations of KMTNet HM events, supplemented by additional observations from the Microlensing Follow-Up Network ($\mu$FUN; \citealt{mufun}) and the KMTNet ``auto-followup'' system. To date, these data have contributed to the analysis of about 20 confirmed or candidate planets, including five systems with $q < 10^{-4}$ \citep{KB200414, KB210171, KB210912, KB220440, KB231431} and a new solution in the ``Planet/Binary'' degeneracy \citep{Three_planet_candidates}.

In this paper, we present two additional low-$q$ planets from the 2025 LCOGT follow-up program: KMT-2025-BLG-0811Lb with $q \sim 4.5\times10^{-5}$ and KMT-2025-BLG-0912Lb, with $q = (2.60 \pm 0.05)\times10^{-4}$. The paper is structured as follows. In Section~\ref{obser}, we describe the observations and data reduction process. In Section~\ref{model}, we present the light-curve modeling. Section~\ref{lens} details the color-magnitude diagram (CMD) analysis and Bayesian analysis to estimate the source and lens properties, respectively. We investigate nine events with ``central-resonant'' degeneracy in Section~\ref{dis}.

\section{Observations}\label{obser}
\begin{figure}
    \centering
    \includegraphics[width=1.0\linewidth]{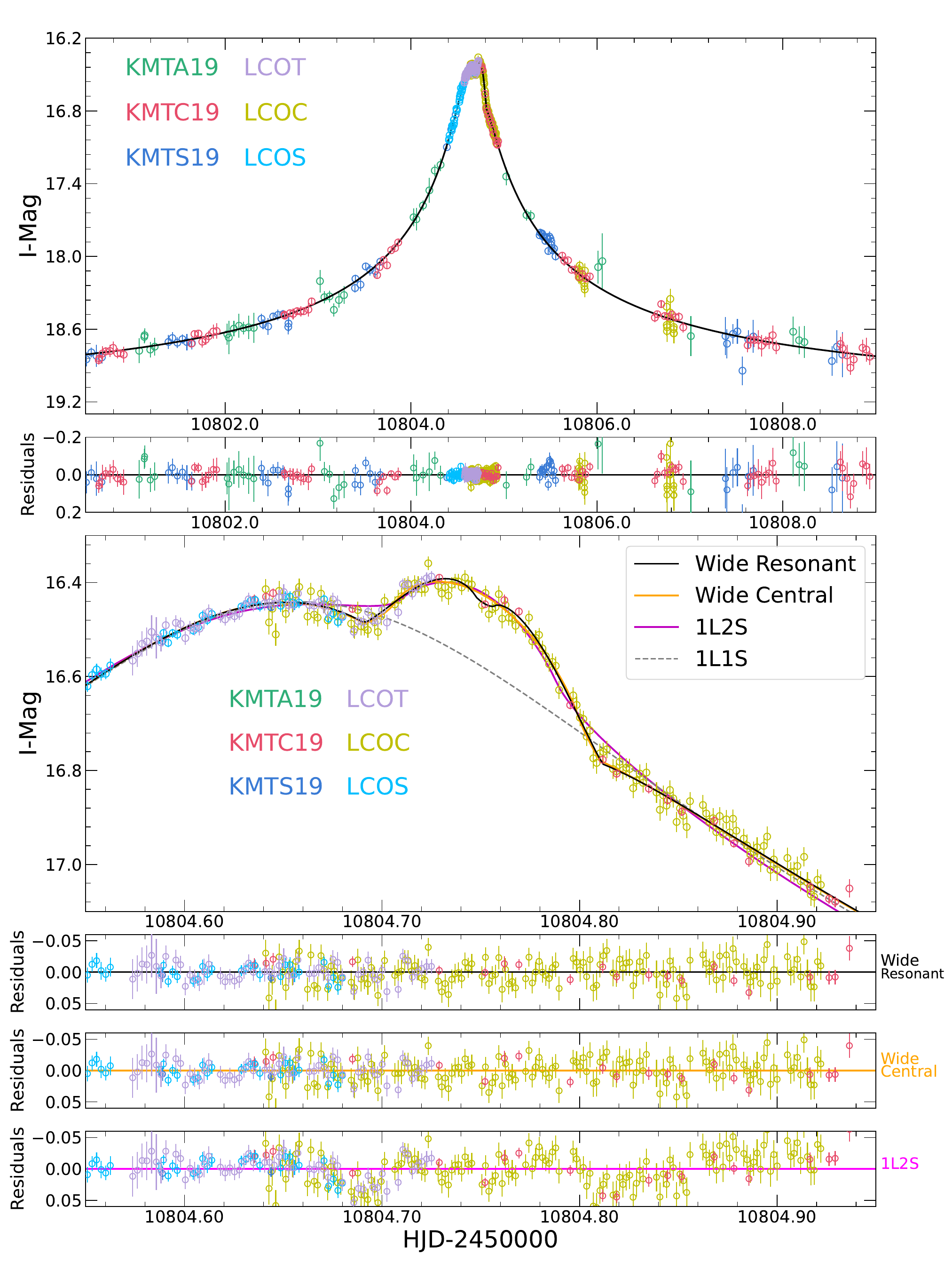}
    \caption{Light curve of the microlensing event, \eventa, with the 2L1S models (solid black and orange lines), 1L2S model (magenta line), and the underlying 1L1S model (dashed grey line). Different data sets are plotted in different colors. The upper panels show the 8-day time interval around the peak. The lower panels present a close-up of the planetary anomaly and the residuals relative to the 2L1S and 1L2S models.}
    \label{fig:250811-lc}
\end{figure}

\begin{figure}
    \centering
    \includegraphics[width=1.0\linewidth]{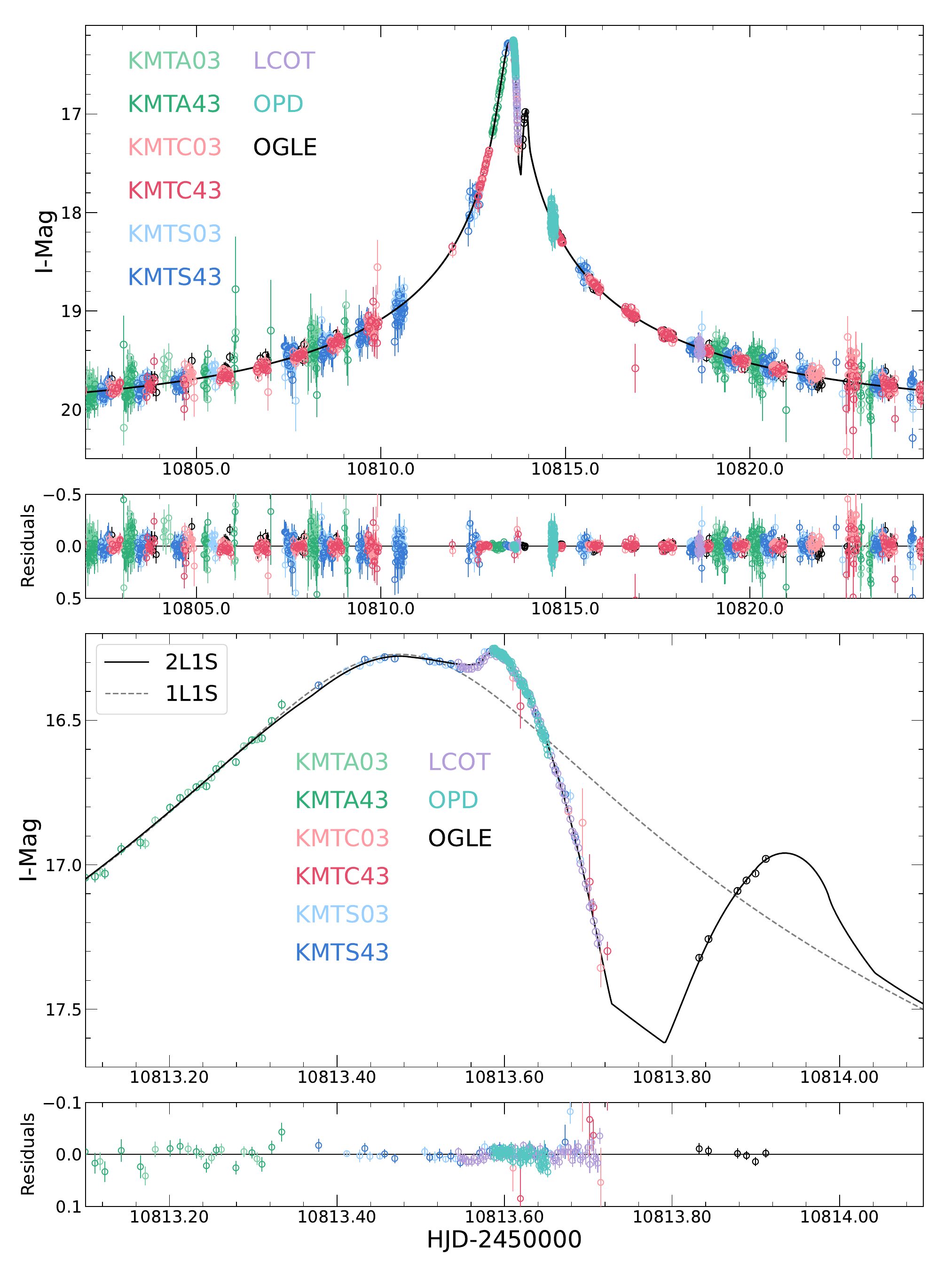}
    \caption{Light curve of the microlensing event, \eventb. The symbols are similar to those in Figure~\ref{fig:250811-lc}, with the 2L1S models (solid black lines) and the underlying 1L1S model (dashed grey line).}
    \label{fig:250912-lc}
\end{figure}

\subsection{Survey Observations}

Figures \ref{fig:250811-lc} and \ref{fig:250912-lc} display the light curves of \eventa\ and \eventb, which were both first alerted by the KMTNet AlertFinder system \citep{KMTAF}. KMTNet operates three identical 1.6\,m telescopes, each equipped with a 4\,deg$^2$ field-of-view camera, located at the Cerro Tololo Inter-American Observatory (CTIO) in Chile (KMTC), the South African Astronomical Observatory (SAAO) in South Africa (KMTS), and the Siding Spring Observatory (SSO) in Australia (KMTA). This global distribution enables nearly continuous monitoring of the Galactic bulge during the bulge season. 

The first alert date of KMT-2025-BLG-0811 is 2025 May 06 at 00:53 (HJD$' \sim 10801.54$, where $\text{HJD}' = \text{HJD} - 2450000$). It is located in the KMTNet BLG19 field at equatorial coordinates $(\alpha, \delta)_{\text{J2000}}$ = (17:42:55.87, $-$24:29:09.20), corresponding to Galactic coordinates $(\ell, b)$ = (3.48, 2.84). The monitoring cadence for this field is $1.0\,\text{hr}^{-1}$. 

KMT-2025-BLG-0912 was identified by the KMTNet AlertFinder on 2025 May 12 at 02:48 ($\text{HJD}' \sim 10807.62$). It is situated at $(\alpha, \delta)_{\text{J2000}}$ = (17:59:23.33, $-$28:48:43.09), or Galactic coordinates $(\ell, b)$ = (1.64, $-$2.53). This event lies in the overlapping region of the BLG03 and BLG43 fields, which are monitored with a combined cadence of $4.0\,\text{hr}^{-1}$. This event was subsequently identified by the Early Warning System \citep{Udalski1994,Udalski2003} of the Optical Gravitational Lensing Experiment \citep[OGLE;][]{OGLEIV} as OGLE-2025-BLG-0653 on 2025 May 16. OGLE observations were conducted using the 1.3\,m Warsaw telescope at Las Campanas Observatory in Chile with a cadence of approximately $1.0\,\text{hr}^{-1}$.

For both events, KMTNet and OGLE images were primarily taken in the $I$ band, supplemented by occasional $V$-band images for source star characterization.

\subsection{Follow-up Observations}

In addition to the survey data, high-cadence follow-up observations were conducted for both events, triggered by the KMTNet \texttt{HighMagFinder} system \citep{KB210171}. This system identifies candidate HM events by monitoring magnification trends in real time using KMTNet photometry.

For \eventa, \texttt{HighMagFinder} issued an alert on 2025 May 06 at 06:15 UT ($\text{HJD}' = 10801.76$). Following the alert, the MAP collaboration initiated follow-up observations using the 1.0\,m telescopes of the LCOGT at three southern sites: CTIO (LCOC), SAAO (LCOS), and Teide Observatory in Spain (LCOT). Furthermore, the KMTNet ``auto-followup'' system was activated from $\text{HJD}' = 10804.76$ to $\text{HJD}' = 10804.94$. During this period, observations of the BLG43 fields ($\Gamma = 2\,\text{hr}^{-1}$) were suspended and replaced with monitoring of the BLG19 field to ensure dense coverage of the anomaly.

\eventb\ was flagged by \texttt{HighMagFinder} on 2025 May 17 at 03:15~UT ($\text{HJD}' = 10812.64$). The first half of the anomaly was covered by three observing components: monitoring with LCOT, follow-up observations by $\mu$FUN using the 0.6\,m telescope at the Observat\'orio do Pico dos Dias (OPD) in Brazil, and survey data from KMTNet. In addition, OGLE survey observations obtained between $\text{HJD}' = 10813.8$ and 10813.9 captured the second half of the anomaly.

\subsection{Data Reduction and Error Analysis}

The photometry of \eventa\ and \eventb\ was conducted using the Difference Image Analysis (DIA) technique \citep{Tomaney1996, Alard1998}. We used the \texttt{pySIS} pipeline \citep{pysis, Yang_TLC,Yang_TLC2} for KMTNet, LCOGT, and $\mu$FUN data, and the \citet{Wozniak2000} pipeline for OGLE. 
To ensure a consistent photometric system for CMD analysis, the $V$ and $I$-band magnitudes were calibrated to the OGLE-III standard scale \citep{OGLEIII}. The error bars from each DIA pipeline are re-normalized following the procedure of \cite{MB11293}, which adjusts the error bars to ensure that the $\chi^2$ per degree of freedom (dof) for each data set is unity. Table~\ref{tab:reduction} summarizes the observation and data reduction information.

\begin{deluxetable*}{l l c c c c c}
\tablecaption{Data information and corresponding data reduction methods
\label{tab:reduction}}
\tablewidth{0pt}
\setlength{\tabcolsep}{12pt} 
\renewcommand{\arraystretch}{1.15}
\tablehead{
\colhead{Event} &
\colhead{Collaboration} &
\colhead{Site} &
\colhead{Name} &
\colhead{Filter$^{1}$} &
\colhead{$N_{\rm data}$} &
\colhead{Reduction method} 
}
\startdata
\multirow{7}{*}{KB250811$^2$}
& KMTNet & SSO  & KMTA19 & $I$ & 381 & pySIS$^{3}$  \\
& KMTNet & CTIO & KMTC19 & $I$ & 960 & pySIS         \\
& KMTNet & CTIO & KMTC19 & $V$ & 93 & pySIS         \\
& KMTNet & SAAO & KMTS19 & $I$ &  412 & pySIS        \\
& MAP    & CTIO & LCOC   & $I$ &  167 & pySIS        \\
& MAP    & SAAO & LCOS   & $I$ &  72 & pySIS         \\
& MAP    & Teide Observatory  & LCOT  & $I$ &  86 & pySIS   \\
\hline
\multirow{10}{*}{KB250912}
& KMTNet & SSO  & KMTA03 & $I$ &  1711 & pySIS  \\
& KMTNet & SSO  & KMTA43 & $I$ &  1753 & pySIS  \\
& KMTNet & CTIO & KMTC03 & $I$ &  2176 & pySIS    \\
& KMTNet & CTIO & KMTC03 & $V$ &  39 & pySIS    \\
& KMTNet & CTIO & KMTC43 & $I$ &  1614 & pySIS    \\
& KMTNet & SAAO & KMTS03 & $I$ &  1828 & pySIS     \\
& KMTNet & SAAO & KMTS43 & $I$ &  1858 & pySIS   \\
& OGLE & Las Campanas Observatory & OGLE & $I$ & 578 & \citet{Wozniak2000}  \\
& MAP    & Teide Observatory  & LCOT   & $I$ &  78 & pySIS       \\
& $\mu$FUN & Observat\'orio do Pico dos Dias & OPD     & $I$ & 171 & pySIS \\
\enddata 
\textbf{Notes.}\\
$^{1}$ Event names are abbreviations, e.g., KMT-2025-BLG-0811 to KB250811.\\
$^{2}$ The V-band data are used only for CMD analysis and are not included in the light-curve modeling.\\
$^{3}$ \citet{pysis, Yang_TLC,Yang_TLC2}.
\end{deluxetable*}

\section{Light-curve Analysis}\label{model}

Figures~\ref{fig:250811-lc} and \ref{fig:250912-lc} display the observed data together with various models for \eventa\ and \eventb, respectively. For both events, the observed light curves exhibit significant deviations from the standard single-lens single-source (1L1S) model.

For \eventa, a short-lived bump ($\sim 2.4$\,hr) is clearly observed shortly after the peak of the 1L1S model, which was captured by the KMTC19, LCOT, and LCOC data. In addition, the LCOS data partially cover the dip before the bump. Such an anomaly could be produced by either a binary-lens single-source (2L1S) model or a single-lens binary-source (1L2S) model \citep{Gaudi1998}. While both models are initially considered, the 1L2S model is later excluded based on the analysis below. For \eventb, the anomaly consists of two distinct bumps separated by a dip. The first bump was well-captured by the KMTNet, LCOT, and OPD data, and the second was only covered by OGLE. Because the morphology of such an anomaly cannot be produced by a 1L2S model, we only consider the 2L1S model for this event.

A standard 2L1S model requires seven parameters to describe the magnification $A(t)$. The first three are the Paczy\'{n}ski parameters $(t_0, u_0, \te)$ \citep{Paczynski1986}, where $t_0$ is the time of the closest approach of the source to the lens center of mass, $u_0$ is the impact parameter normalized to the angular Einstein radius $\thetae$, and $\te$ is the Einstein radius crossing time, defined as:
\begin{equation}
\te = \frac{\thetae}{\mu_{\rm rel}}; \quad \thetae = \sqrt{\kappa M_{\rm L} \pi_{\rm rel}},
\end{equation}
where $M_{\rm L}$ is the lens mass, $(\pi_{\rm rel}, \mu_{\rm rel})$ are the lens-source relative parallax and proper motion, and $\kappa \equiv 4G/(c^2\mathrm{au}) \simeq 8.144~\mathrm{mas}\,M_\odot^{-1}$. The next three parameters $(q, s, \alpha)$ define the binary-lens geometry, representing the binary mass ratio $q$, the projected separation $s$ normalized to $\thetae$, and the angle $\alpha$ between the source trajectory and the binary axis, respectively. The final parameter is the normalized source radius $\rho = \theta_*/\thetae$, where $\theta_*$ is the angular source radius.

For each data set $i$, we introduce two flux parameters, $f_{{\rm S},i}$ and $f_{{\rm B},i}$, representing the flux of the source star and any blended flux (including the lens flux and light from nearby stars). The observed flux $f_i(t)$ is modeled as:
\begin{equation}
f_i(t) = f_{{\rm S},i}\,A(t) + f_{{\rm B},i},
\end{equation}
where $A(t)$ is the 2L1S magnification computed using the contour-integration code \texttt{VBBinaryLensing} \citep{Bozza2010,Bozza2018,VBMicrolensing2025}.

To systematically explore the 2L1S parameter space and identify all local $\chi^2$ minima, we adopt a two-stage grid-search strategy. We first conduct a sparse grid over $(\log s, \log q, \log \rho, \alpha)$, sampling 61 values in $-1.5 \le \log s \le 1.5$, 61 values in $-6 \le \log q \le 0$, 9 values in $-4.0 \le \log \rho \le -1.6$, and 16 trial angles uniformly distributed in $0^\circ \le \alpha < 360^\circ$. Based on the results of the sparse grid, we then perform a denser grid search over a narrower parameter range. At each grid point, we perform a $\chi^2$ minimization using the Markov chain Monte Carlo (MCMC) \texttt{emcee} ensemble sampler \citep{emcee}, holding $(\log s, \log q, \log \rho)$ fixed while allowing $(t_0, u_0, \te, \alpha)$ to vary. After identifying potential local minima, we refine each solution by allowing all seven parameters to vary in an MCMC.

For the standard 1L2S model, the light curve is the superposition of two 1L1S curves, requiring eight parameters \citep{MB12486}: $(t_{0,1}, u_{0,1}, \rho_1)$ and $(t_{0,2}, u_{0,2}, \rho_2)$ describe the impact time, impact parameter, and normalized size for the two sources, respectively, while $\te$ is assumed to be identical for both. The flux ratio of the two sources is defined as $q_f = f_{{\rm S},2}/f_{{\rm S},1}$, where ``1'' and ``2'' represent the primary and the secondary sources, respectively.

We also examine the possible contribution of the microlensing parallax effect caused by the orbital motion of Earth \citep{Gould1992,Gould2000}. The microlensing parallax vector is defined as
\begin{equation}
\boldsymbol{\pi}_{\rm E} = \frac{\pi_{\rm rel}}{\theta_{\rm E}} \frac{\boldsymbol{\mu}_{\rm rel}}{\mu_{\rm rel}}.
\end{equation}
We fit it by including two parameters, $\pi_{\rm E,N}$ and $\pi_{\rm E,E}$, the north and east components of the microlensing parallax vector in equatorial coordinates. We also account for the ecliptic degeneracy \citep{OB09020} by fitting the $u_0 > 0$ and $u_0 < 0$ solutions.

If finite-source effects are present during caustic crossings or cusp approaches, we include limb-darkening effects. We adopt a linear limb-darkening law \citep{An2002,Claret2011},
\begin{equation}
S_\lambda(\mu) = S_\lambda(1)\,[1 - u_\lambda(1-\mu)],
\end{equation}
where $S_\lambda(1)$ is the surface brightness at the center of the source, $\mu$ is the cosine of the angle between the line of sight and the local surface normal, and $u_\lambda$ is the limb-darkening coefficient in band $\lambda$. For each event, the limb-darkening coefficients are inferred from the estimated effective temperature $T_{\rm eff}$ of the source star using the tabulations of \citet{Claret2011}.

\subsection{KMT-2025-BLG-0811}

\begin{figure}
    \includegraphics[width=0.47\textwidth]{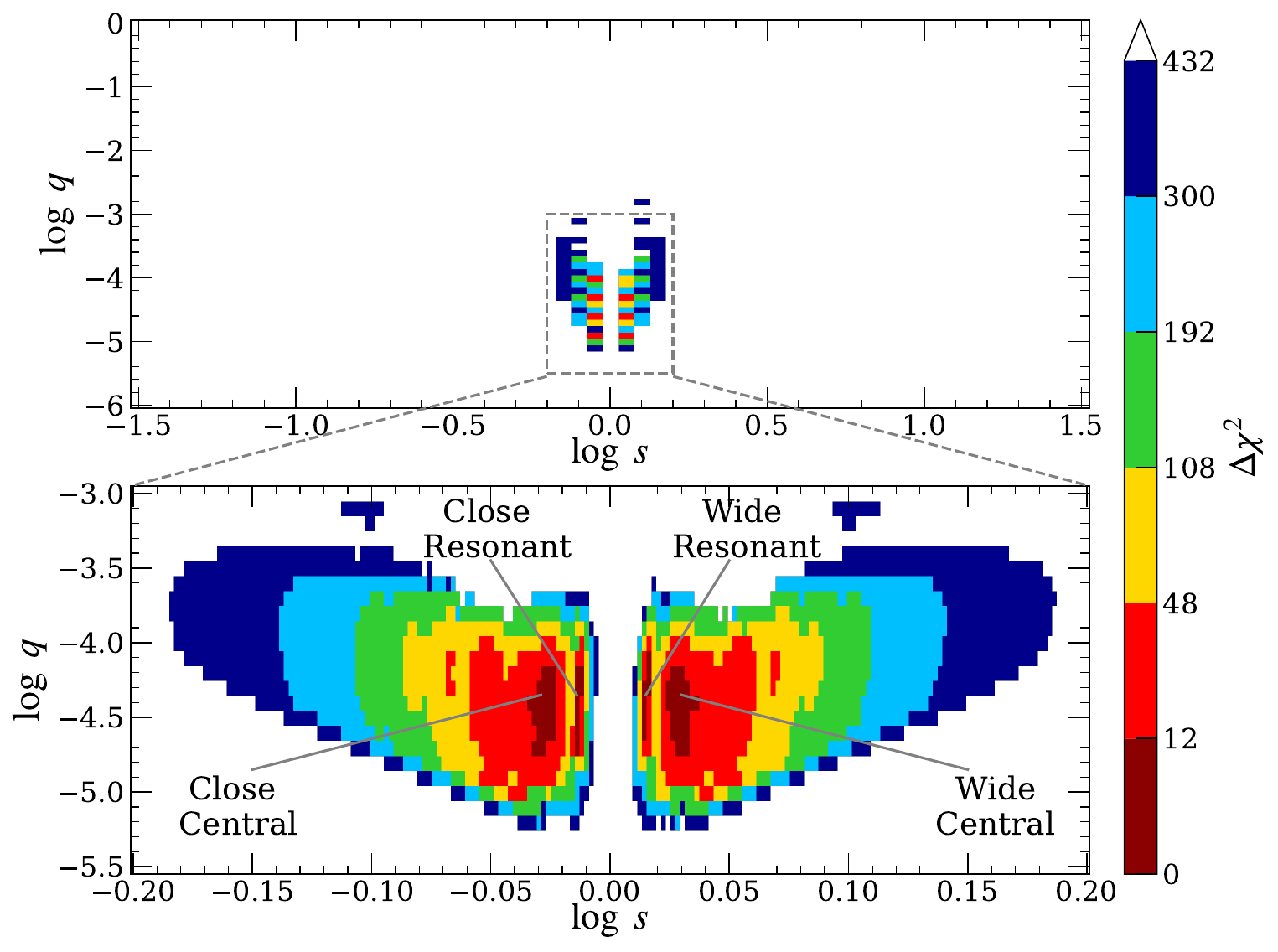}
    \caption{The $\chi^2$ landscape in the $(\log s, \log q)$ parameter space derived from the 2L1S grid search for \eventa. The top panel presents the results from the initial coarse grid, while the bottom panel provides a magnified view of the refined grid in the vicinity of the global minimum. Color shading from dark red to dark blue corresponds to regions within $<1n\sigma$, $<2n\sigma$, $<3n\sigma$, $<4n\sigma$, $<5n\sigma$, and $<6n\sigma$, respectively, where $n^2 = 12$. Grid points beyond the $6n\sigma$ threshold are not shown. }
\label{fig:grid-0811}
\end{figure}

From the CMD analysis in Section~\ref{CMD}, we infer an effective temperature of $T_{\rm eff}\simeq4689$~K for the source star. We therefore adopt a linear limb-darkening coefficient $u_I=0.59$ for the $I$ band from \citet{Claret2011}. We conduct a systematic grid search for 2L1S solutions as described in the previous section. Figure~\ref{fig:grid-0811} displays the $\chi^2$ distribution in the $(\log s, \log q)$ plane from the initial search. The upper panel reveals distinct minima within the regions $-0.2 \le \log s \le 0.2$, $-5.5 \le \log q \le -3.0$, and $-4.0 \le \log \rho \le -2.5$. Consequently, we perform a dense follow-up grid search comprising 201 values in $-0.2\le\log s\le0.2$, 26 values in $-5.5\le\log q\le-3.0$, 16 values in $-4.0\le\log\rho\le-2.5$, and 16 initial source-trajectory angles uniformly distributed over $0^\circ \le \alpha < 360^\circ$. The dense grid search yields two pairs of local minima, following the well-known ``close-wide'' degeneracy \citep{Griest1998} and the ``central-resonant'' degeneracy (e.g., \citealt{KMT2021_mass1,KB210171}). Following the standard nomenclature, we label these four solutions as ``close central'', ``wide central'', ``close resonant'', and ``wide resonant''.

\begin{table*}[htb]
\centering
\small
\setlength{\tabcolsep}{6pt}
\renewcommand{\arraystretch}{1.2}
\caption{2L1S Lensing Parameters}
\begin{tabular}{l | cc | cc | c}
\hline\hline

& \multicolumn{4}{c|}{\textbf{\eventa}}
& \textbf{\eventb} \\
Parameter
& Close Central & Wide Central
& Close Resonant & Wide Resonant
&  \\

\hline

$\chi^2/{\rm dof}$
& 2082.7/2071 & 2082.8/2071
& 2075.8/2071 & 2074.6/2071
& 11771.2/11760 \\

\hline

$t_0$ (HJD$^\prime$)
& $10804.6649 \pm 0.0008$ 
& $10804.6650 \pm 0.0008$
& $10804.6652 \pm 0.0008$
& $10804.6651 \pm 0.0008$
& $10813.4682 \pm 0.0010$ \\

$u_0$
& $0.00319 \pm 0.00047$
& $0.00317 \pm 0.00046$
& $0.00320 \pm 0.00048$
& $0.00328 \pm 0.00050$
& $0.01587 \pm 0.00021$ \\

$t_{\rm E}$ (day)
& $50.05 \pm 7.33$
& $50.53 \pm 7.28$
& $49.82 \pm 7.57$
& $48.55 \pm 7.62$
& $13.65 \pm 0.13$ \\

$\rho$ ($10^{-3}$)
& $0.93 \pm 0.13$
& $0.92 \pm 0.13$
& $0.48 \pm 0.07$
& $0.50 \pm 0.08$
& $3.87 \pm 0.05$ \\

$\alpha$ (deg)
& $298.75 \pm 0.24$
& $298.71 \pm 0.25$
& $298.86 \pm 0.25$
& $298.79 \pm 0.23$
& $143.03 \pm 0.19$ \\

$s$
& $0.932 \pm 0.008$
& $1.076 \pm 0.009$
& $0.970 \pm 0.001$
& $1.035 \pm 0.001$
& $0.9948 \pm 0.0012$ \\

$q$ ($10^{-5}$)
& $4.397 \pm 0.681$
& $4.356 \pm 0.650$
& $4.522 \pm 0.685$
& $4.663 \pm 0.724$
& $25.974 \pm 0.468$ \\

$I_{\rm S,OGLE}$
& $22.579 \pm 0.159$
& $22.589 \pm 0.156$
& $22.570 \pm 0.162$
& $22.543 \pm 0.165$
& $20.844 \pm 0.013$ \\
\hline\hline
\end{tabular}
\begin{flushleft}
\textbf{Notes.} 
HJD$^\prime = \mathrm{HJD} -2450000$.
\end{flushleft}
\label{tab:static-parm}
\end{table*}

\begin{figure}
    \includegraphics[width=0.5\textwidth]{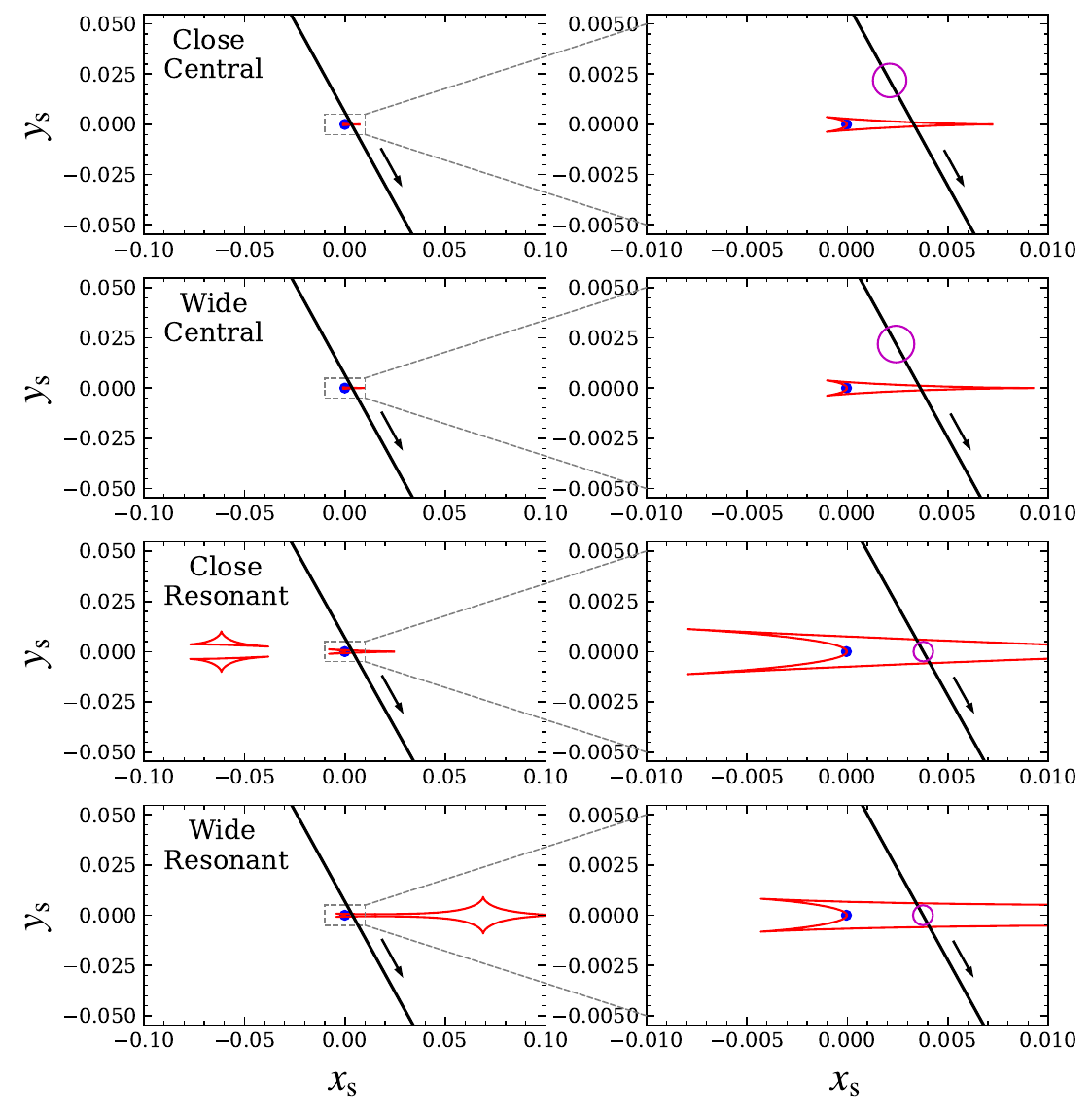}
    \caption{Caustic geometries of the four 2L1S solutions for \eventa. In each panel, the red lines show the caustic, the blue dot marks the location of the host star, the black line represents the source–lens relative trajectory, and the arrow indicates the direction of source motion. The right panels provide close-up views of the caustic-crossing regions, with the radii of the magenta circles indicating the source size.}
\label{fig:cau-250811}
\end{figure}

Each local minimum is further refined by an MCMC process with all parameters set free. The best-fitting parameters for each solution are presented in Table~\ref{tab:static-parm}. Our analysis indicates that the ``wide resonant'' solution provides the best fit to the data. However, the other three solutions, ``close central'', ``wide central'', and ``close resonant'', are only slightly disfavored by $\Delta\chi^2 = 8.1$, $8.2$, and $1.2$, respectively. We note that the ``central'' solutions require a larger source size, with $\rho$ nearly twice that of the ``resonant'' solutions, while yielding similar mass ratios $q$. We will discuss the ``central-resonant'' degeneracy in more detail in Section~\ref{dis}. The model light curves and the corresponding caustic structures are shown in Figure~\ref{fig:250811-lc} and Figure~\ref{fig:cau-250811}. As illustrated, all four solutions involve caustic-crossing trajectories.

\begin{table}[htb]
    \renewcommand\arraystretch{1.20}
    \centering
    \caption{1L2S Parameters for \eventa}
    \begin{tabular}{c|c c}
    \hline
    \hline
    $\chi^2$/dof & $2371.7 / 2078$  \\
    \hline
    $t_{0,1}-10804$ (${\rm HJD}^{\prime}$)  & $0.6488 \pm 0.0009$ \\
    $u_{0,1}$  & $0.00303 \pm 0.00045$ \\
    $t_{0,2}-10804$ (${\rm HJD}^{\prime}$)  & $0.7491 \pm 0.0006$ \\
    $u_{0,2}$  & $0.00001 \pm 0.00031$ \\
    $\rho_2$ ($10^{-3}$) & $0.71 \pm 0.33$ \\
    $\te$ (days)  & $55.45 \pm 5.95$ \\
    $q_f$ & $0.0222 \pm 0.0007$ \\
    $I_{\rm S,OGLE}$ & $22.695 \pm 0.195$ \\
    \hline
    \hline
    \end{tabular}
    \label{tab:0811-1L2S}
\end{table}

We also explore the 1L2S interpretation for \eventa, but this model is strongly disfavored, with $\Delta\chi^2 \sim 297.1$ relative to the preferred 2L1S ``wide resonant'' solution. The resulting parameters are given in Table \ref{tab:0811-1L2S}. As shown in Figure \ref{fig:250811-lc}, the 1L2S model fails to reproduce the bump-like anomaly. The finite-source effect of the secondary source is measured, with $\rho_2 = (0.71 \pm 0.33)\times 10^{-3}$, which yields a lens-source relative proper motion of  $\mu_{\rm rel} \sim 0.49~\mathrm{mas~yr^{-1}}$ according to the CMD analysis in Section \ref{lens}. For typical microlensing events, the cumulative distribution of relative proper motions at low $\mu_{\rm rel}$ can be approximated by \citep{MASADA,2019_subprime}
\begin{equation}
p(\le \mu_{\rm rel}) \simeq 2.8 \times 10^{-2}
\left(\frac{\mu_{\rm rel}}{1~\mathrm{mas~yr^{-1}}}\right)^2 .
\end{equation}
For this event, the probability of obtaining such a slow relative motion is therefore only $\sim 0.67\%$, which further disfavors the 1L2S interpretation. We thus exclude the 1L2S model. 

Due to the very faint source star, $I_{\rm S} \sim 22.6$ magnitude, the inclusion of the parallax effect improves the fit by $\Delta\chi^2 < 1$, and the 1$\sigma$ uncertainties of the parallax vector in all directions are $>0.30$, providing no meaningful constraint. We therefore adopt the static model as the final interpretation of this event.

This event reveals a new low-$q$ microlensing planet with a mass ratio of $q \sim 4.5 \times 10^{-5}$, comparable to the Uranus-to-Sun mass ratio.

\subsection{KMT-2025-BLG-0912}

\begin{figure}
    \includegraphics[width=0.47\textwidth]{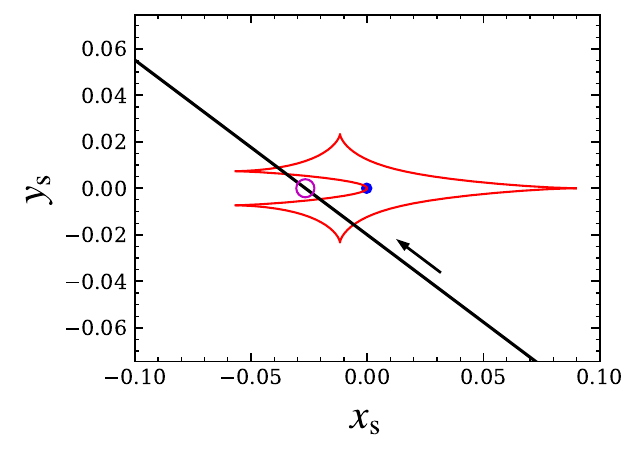}
    \caption{Caustic geometries of the 2L1S solution for \eventb. The symbols are similar to those in Figure~\ref{fig:cau-250811}.}
\label{fig:cau-250912}
\end{figure}

Following the same strategy as for the first event, we conduct a systematic grid search to explore the 2L1S parameter space for \eventb. The grid search yields only a single viable local minimum. Figure~\ref{fig:cau-250912} illustrates the caustic geometry and the source trajectory for the best-fit model. As anticipated from the light curve morphology, the source trajectory crosses a resonant caustic. The first bump and the subsequent peak in the light curve are produced by the source entering and exiting the caustic structure, while the trough between them arises from the relatively demagnified region that is flanked by the ``back end'' of caustic structure (opposite the planet).

The 2L1S parameters derived from the MCMC refinement are presented in Table~\ref{tab:static-parm}. Despite some gaps in the observational coverage of the anomaly, the finite-source effects are well-characterized, with a measured normalized source radius of $\rho = (3.87 \pm 0.05) \times 10^{-3}$. 

We also examine the microlensing parallax effect, but due to the short event timescale, the parallax signal is weak and does not lead to useful constraints, with the $1\sigma$ uncertainty of $\pie$ exceeding 0.5 for all directions. We therefore adopt the static model for this event. 

With $q = 2.6 \times 10^{-4}$, this event has a planet with a mass ratio close to that of Saturn relative to the Sun.

\section{Source and Lens Properties}\label{lens}

\begin{figure}[htb] 
    \centering
    \includegraphics[width=0.45\textwidth]{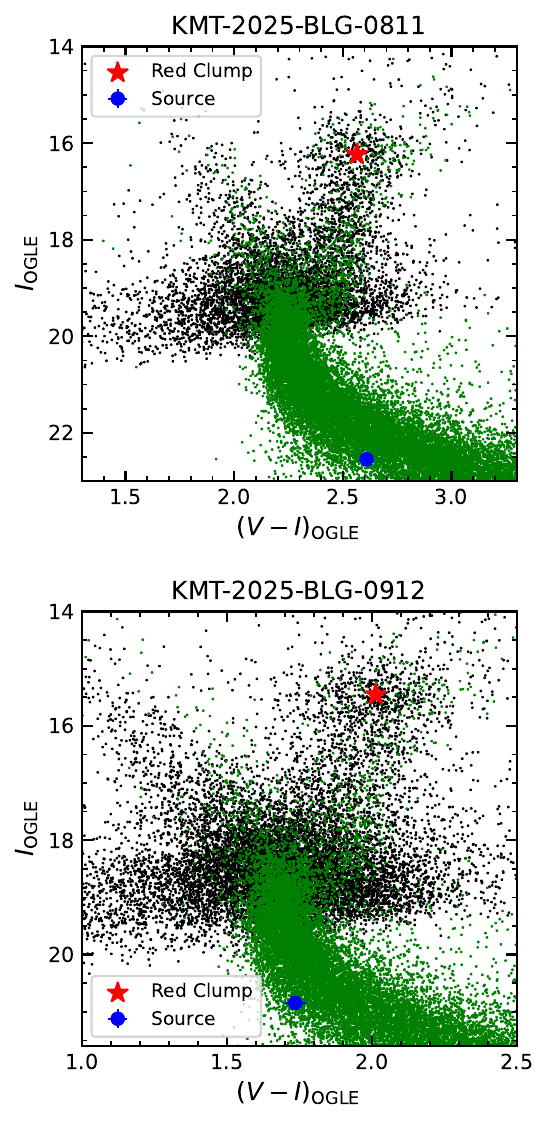}
    \caption{Color-magnitude diagrams for \eventa\ (upper panel) and \eventb\ (lower panel). Black dots show OGLE-III field stars \citep{OGLEIII} within $2'$ of the event positions. The red asterisk and blue dot mark the centroids of the red clump and the source stars, respectively. Green points show the \textit{HST} CMD from \cite{HSTCMD}, whose red clump has been shifted to match the OGLE-III red clump using $(V - I, I)_{\rm RC,HST} = (1.62, 15.15)$ \citep{MB07192}. Both sources lie near the blue edge of the bulge main sequence, suggesting relatively low metallicity.}
    \label{fig:cmd}
\end{figure}

\subsection{Color-Magnitude Diagram}\label{CMD}
The purpose of this section is to derive the intrinsic properties of the source stars by placing them on a $V - I$ versus $I$ CMD, and then to estimate the angular source radius $\theta_\ast$ \citep{Yoo2004}. Combined with the normalized source size $\rho$ from the light-curve modeling, this yields the angular Einstein radius and the lens-source relative proper motion,
\begin{equation}
\theta_{\rm E} = \frac{\theta_\ast}{\rho}, \qquad
\mu_{\rm rel} = \frac{\theta_{\rm E}}{t_{\rm E}} .
\label{eq:thetaE_murel}
\end{equation}

For each event, we construct a $V-I$ versus $I$ CMD using stars from the OGLE-III catalog within a $2'$ radius centered on the event position \citep{OGLEIII}. Figure~\ref{fig:cmd} shows the CMDs for both events.
The centroid of the red clump (RC) is identified following the method of \citet{Nataf2013}. Because RC giants act as approximate standard candles, the offset between the observed RC centroid and the intrinsic RC position provides an estimate of the line-of-sight extinction and reddening. This calibration is then applied to the source star to derive its intrinsic color and magnitude.

The apparent source color $(V-I)_{\rm S}$ is measured from a regression of the KMTC $V$-band and $I$-band fluxes, while the apparent source magnitude $I_{\rm S}$ is determined from the light-curve modeling. After calibration to the OGLE-III photometric system, the source position is located on the CMD. 
Using the offset between the RC centroid and the source position, 
\begin{equation}
    \Delta[(V-I), I] = [(V-I), I]_{\rm S} - [(V-I), I]_{\rm RC},
\end{equation} 
the intrinsic source properties are calculated as 
\begin{equation}
    [(V-I), I]_{\rm S,0} = [(V-I), I]_{\rm RC,0} + \Delta[(V-I), I],
\end{equation}
where the de-reddened color and magnitude of RC, $[(V-I), I]_{\rm RC,0}$, are from \cite{Bensby2013} and Table 1 of \cite{Nataf2013}. Finally, the angular source radius $\theta_\ast$ is estimated using the color/surface-brightness (CSB) relation of \citet{Adams2018},
\begin{equation}
\log(2\theta_\ast) = 0.378 (V-I)_{\rm S,0} + 0.542 - 0.2 I_{\rm S,0},
\end{equation}
which leads to measurements of $\theta_{\rm E}$ and $\mu_{\rm rel}$ through Equation~(\ref{eq:thetaE_murel}). We also estimate $T_{\rm eff}$ through the color-temperature relations of \cite{Houdashelt2000} to determine the limb-darkening coefficients used in Section \ref{model}.

The relative CMD parameters, derived source properties, $\theta_{\rm E}$, and $\mu_{\rm rel}$ for all solutions are summarized in Table~\ref{tab:cmd}. 
Based on the color-spectral type relations of \citet{Bessell1988}, the source star of \eventa\ is most consistent with a K-type dwarf, while that of \eventb\ is likely a G-type dwarf.

For \eventa, the ``central'' and ``resonant'' solutions predict substantially different relative proper motions: 
$\mu_{\rm rel} \sim 2.6$~mas~yr$^{-1}$ for the ``central'' solutions and 
$\mu_{\rm rel} \sim 5.1$~mas~yr$^{-1}$ for the ``resonant'' solutions. 
This factor of $\sim2$ difference implies markedly different lens-source separations at future epochs. Therefore, the ``central-resonant'' degeneracy in \eventa\ could be resolved in the future through high-resolution imaging by separately resolving the lens and source and thereby directly measuring $\mu_{\rm rel}$.

\begin{table*}[htb]
\centering
\small
\setlength{\tabcolsep}{6pt}
\renewcommand{\arraystretch}{1.2}
\caption{CMD parameters, source properties and derived $\theta_{\rm E}$ and $\mu_{\rm rel}$}
\begin{tabular}{l | cc | cc | c}
\hline\hline

& \multicolumn{4}{c|}{\textbf{\eventa}}
& \textbf{\eventb} \\
Parameter
& Close Central & Wide Central 
& Close Resonant & Wide Resonant 
&  \\

\hline
$I_{\rm RC}$
& 16.24 $\pm$ 0.05
& $\xleftarrow{}$
& $\xleftarrow{}$
& $\xleftarrow{}$
& 15.46 $\pm$ 0.04\\

$(V-I)_{\rm RC}$
& 2.57 $\pm$ 0.01
& $\xleftarrow{}$
& $\xleftarrow{}$
& $\xleftarrow{}$
& 2.01 $\pm$ 0.01\\

$I_{\rm RC,0}$
& 14.34 $\pm$ 0.04
& $\xleftarrow{}$
& $\xleftarrow{}$
& $\xleftarrow{}$
& 14.38 $\pm$ 0.04\\

$(V-I)_{\rm RC,0}$
& 1.06 $\pm$ 0.03
& $\xleftarrow{}$
& $\xleftarrow{}$
& $\xleftarrow{}$
& $\xleftarrow{}$\\

$I_{\rm S}$
& 22.58 $\pm$ 0.16
& 22.59 $\pm$ 0.16
& 22.57 $\pm$ 0.16
& 22.54 $\pm$ 0.17
& 20.84 $\pm$ 0.01 \\

$(V-I)_{\rm S}$
& 2.61 $\pm$ 0.03
& $\xleftarrow{}$
& $\xleftarrow{}$
& $\xleftarrow{}$
& 1.74 $\pm$ 0.04\\

$I_{\rm S,0}$
& 20.68 $\pm$ 0.17
& 20.69 $\pm$ 0.17
& 20.67 $\pm$ 0.17
& 20.65 $\pm$ 0.18
& 19.77 $\pm$ 0.06 \\

$(V-I)_{\rm S,0}$
& 1.10 $\pm$ 0.04
& $\xleftarrow{}$
& $\xleftarrow{}$
& $\xleftarrow{}$
& 0.78 $\pm$ 0.05\\

$\theta_\ast$ ($\mu$as)
& 0.333 $\pm$ 0.031
& 0.331 $\pm$ 0.031
& 0.334 $\pm$ 0.032
& 0.338 $\pm$ 0.032
& 0.383 $\pm$ 0.021 \\

$\theta_{\rm E}$ (mas)
& 0.359 $\pm$ 0.062
& 0.361 $\pm$ 0.062
& 0.694 $\pm$ 0.125
& 0.683 $\pm$ 0.125
& 0.099 $\pm$ 0.006 \\

$\mu_{\rm rel}$ (mas\,yr$^{-1}$)
& 2.620 $\pm$ 0.594
& 2.605 $\pm$ 0.581
& 5.081 $\pm$ 1.195
& 5.134 $\pm$ 1.236
& 2.643 $\pm$ 0.153 \\

\hline\hline
\end{tabular}
\label{tab:cmd}
\end{table*}

\subsection{Bayesian Analysis}\label{Baye}

\begin{figure*}[htb] 
    \centering
    \includegraphics[width=1.0\textwidth]{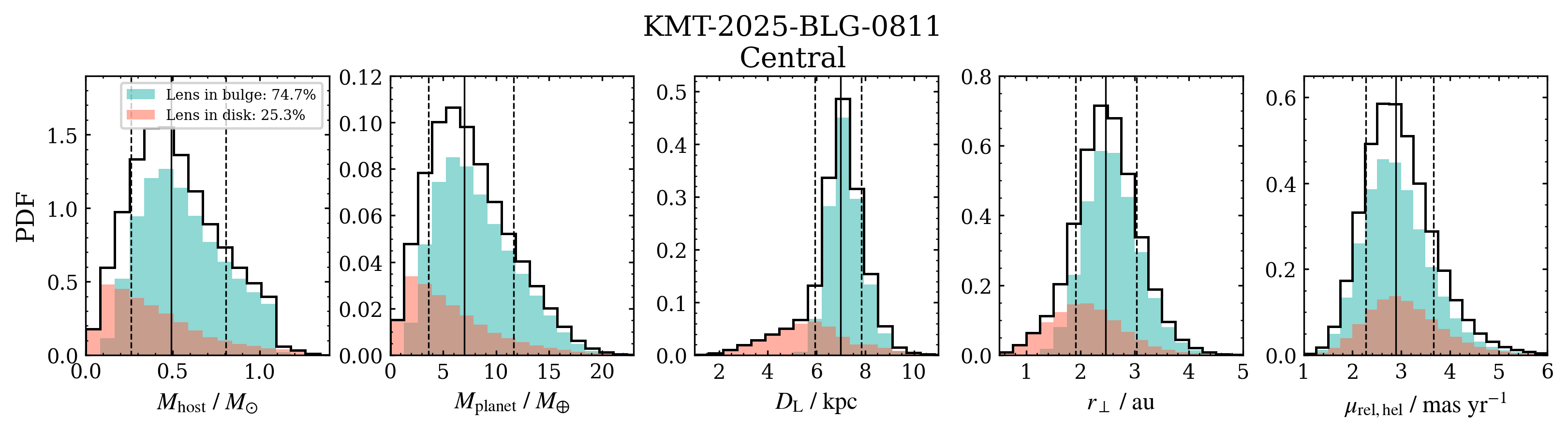}
    \includegraphics[width=1.0\textwidth]{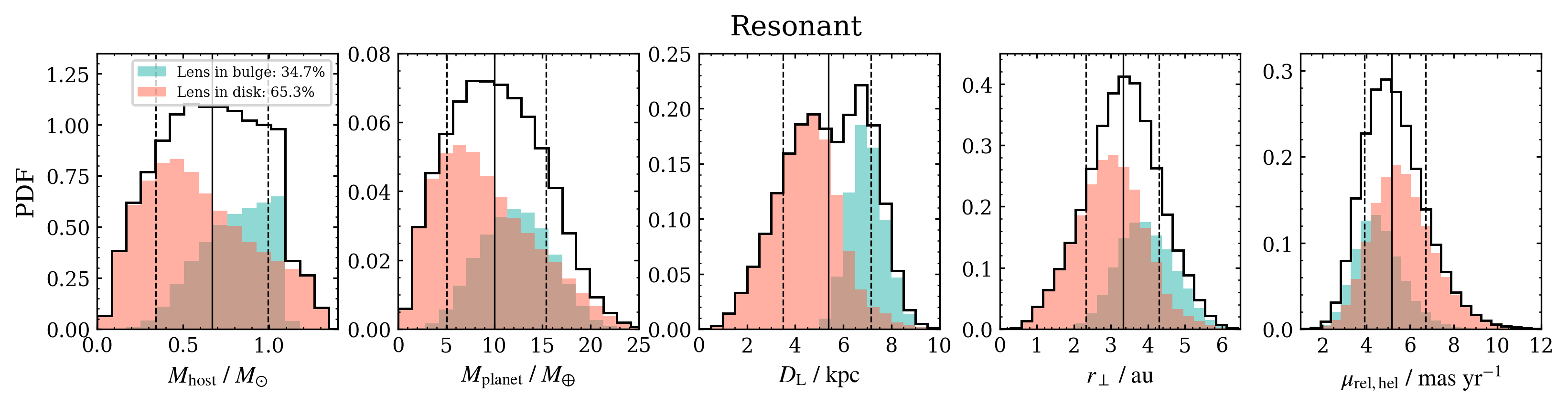}
    \includegraphics[width=1.0\textwidth]{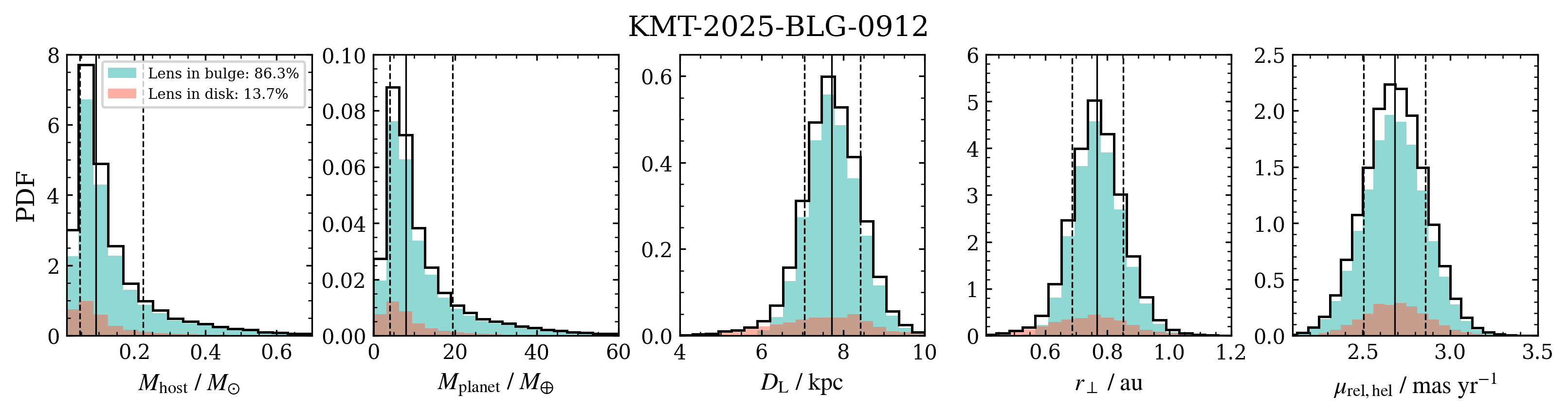}
    \caption{Posterior probability distributions from our Bayesian analysis are shown for the host mass $M_{\rm host}$, the planetary mass $M_{\rm planet}$, the lens distance $D_{\rm L}$, the projected planet-host separation $r_\bot$, and the lens-source relative proper motion in the heliocentric frame $\mu_{\rm rel,hel}$. In each panel, the solid black line indicates the median, and the two dashed black lines mark the 15.9\% and 84.1\% credible limits. Contributions from bulge and disk lens populations are shown in red and green, respectively.}
    \label{fig:baye}
\end{figure*}

If both the angular Einstein radius $\theta_{\rm E}$ and the microlensing parallax vector $\boldsymbol{\pi}_{\rm E}$ are measured, the lens mass $M_{\rm L}$ and distance $D_{\rm L}$ can be directly determined \citep{Gould1992,Gould2000}
\begin{equation}\label{eq:mass}
    M_{\rm L} = \frac{\theta_{\rm E}}{\kappa \pi_{\rm E}}, \qquad 
    D_{\rm L} = \frac{\mathrm{au}}{\pi_{\rm E}\theta_{\rm E} + \mathrm{au}/D_{\rm S}},
\end{equation}
where $\kappa \equiv 4G/(c^2{\rm au}) \simeq 8.144~{\rm mas}~M_\odot^{-1}$ and $D_{\rm S}$ is the source distance.
However, there is no meaningful parallax signal for either \eventa\ or \eventb. We therefore estimate the physical parameters of the lens systems using a Bayesian analysis with Galactic-model priors.

The Galactic model consists of the lens mass function, spatial density distributions, and kinematics for both disk and bulge populations. We adopt the \citet{Kroupa2001} initial mass function with upper-mass cutoffs of $1.3\,M_\odot$ for disk lenses and $1.1\,M_\odot$ for bulge lenses \citep{Zhu2017spitzer}.
The stellar density profiles follow \citet{Yang2021_GalacticModel}.
Bulge kinematics are modeled with zero mean velocity and a one-dimensional velocity dispersion of $120~{\rm km~s^{-1}}$ in each direction \citep{Zhu2017spitzer}, while disk kinematics follow the ``Model C'' prescription of \citet{Yang2021_GalacticModel}, implemented using the \texttt{galpy} package \citep{bovy2015galpy}.

We generate $10^7$ simulated events from the Galactic model and weight them by 
\begin{equation}\label{eq:weights}
    w_i = \Gamma_i \times \mathcal{L}_i(t_{\rm E}) \, \mathcal{L}_i(\theta_{\rm E}) ,
\end{equation}
where $\Gamma_i \propto \theta_{{\rm E},i}\,\mu_{{\rm rel},i}$ is the event rate factor. The likelihood terms $\mathcal{L}_i(t_{\rm E})$$\mathcal{L}_i(\theta_{\rm E})$ are derived from the posterior constraints obtained in our light-curve analysis.
For all solutions, the posteriors of $t_{\rm E}$ and $\theta_{\rm E}$ are well approximated by Gaussian distributions.

\begin{table*}[htb]
\renewcommand\arraystretch{1.25}
\centering
\caption{Lensing Physical Parameters from Bayesian Analyses}
\begin{tabular}{c c | c c c c c}
\hline
\hline
Event & Model 
& $M_\mathrm{host}\,(M_\odot)$ 
& $M_\mathrm{planet}\,(M_\oplus)$ 
& $D_\mathrm{L}\,(\mathrm{kpc})$ 
& $r_\bot\,(\mathrm{au})$ 
& $\mu_\mathrm{rel,hel}\,(\mathrm{mas\,yr^{-1}})$ \\
\hline
\multirow{2}{*}{KMT-2025-BLG-0811}
& Central
& $0.49_{-0.23}^{+0.32}$ 
& $6.99_{-3.35}^{+4.66}$ 
& $6.99_{-1.06}^{+0.85}$ 
& $2.47_{-0.56}^{+0.57}$ 
& $2.89_{-0.61}^{+0.77}$ \\
& Resonant
& $0.67_{-0.33}^{+0.33}$ 
& $10.00_{-4.96}^{+5.37}$ 
& $5.39_{-1.89}^{+1.75}$ 
& $3.34_{-1.01}^{+0.96}$ 
& $5.16_{-1.25}^{+1.55}$ \\
\hline
KMT-2025-BLG-0912
&
& $0.092_{-0.045}^{+0.132}$ 
& $7.98_{-3.91}^{+11.43}$ 
& $7.72_{-0.71}^{+0.66}$ 
& $0.77_{-0.08}^{+0.08}$ 
& $2.68_{-0.18}^{+0.18}$ \\
\hline
\hline
\end{tabular}
\label{tab:baye}
\end{table*}

The physical parameters inferred from the Bayesian analysis are summarized in Table~\ref{tab:baye} and illustrated in Figure~\ref{fig:baye}. We report the host mass $M_{\rm host}$, the planetary mass $M_{\rm planet}$, the lens distance $D_{\rm L}$, the projected planet-host separation  $r_\perp$, and the heliocentric lens-source relative proper motion $\mu_{\rm rel,hel}$. 

For \eventa, the Bayesian posteriors are obtained by averaging over the degenerate ``close'' and ``wide'' solutions in both ``central'' and ``resonant'' configuration.
Each solution is weighted by its relative Galactic-model probability and by ${\rm exp}(-\Delta\chi^2/2)$, where $\Delta\chi^2$ is measured relative to the best-fitting model within each configuration.

The results show that ``central'' and ``resonant'' models for \eventa\ favor an M- or K-dwarf host. The inferred planetary mass falls in the super-Earth/mini-Neptune regime, with projected separations of $r_\perp \sim 2.5$~au for the central topology and $r_\perp \sim 3.3$~au for the resonant topology. Adopting a water snow-line scaling of $a_{\rm SL} = 2.7(M/M_\odot)$~au \citep{snowline}, the planet lies beyond the snow line in both cases. For the central topology, the lens system is more likely located in the Galactic bulge, whereas for the resonant topology a disk lens is favored. Using the color-color relation of \cite{Bessell1988}, the source has an apparent $K$-band magnitude of $K_S \sim 19.5$. For a host with $M_{\rm host} = 0.67 M_\odot$ at 5.4~kpc, the expected lens brightness is $K_{\rm L} \sim 19.1$. Therefore, immediate adaptive optics observations with Keck could detect excess flux from the lens at the source position.

The Bayesian analysis indicates that the companion of \eventb\ is also a super-Earth/mini-Neptune, with a mass of $M_{\rm planet}\sim8\,M_\oplus$ and a projected separation of $r_\perp\sim0.8$~au. The host is inferred to be a very low-mass star, with $M_{\rm host}\sim0.09\,M_\odot$, likely located in the Galactic bulge. For such a low-mass host, the corresponding snow-line distance is $a_{\rm SL}\sim0.25$~au, implying that the planet is also located beyond the snow line.

\section{Discussion: Central-Resonant Degeneracy}\label{dis}

\begin{table*}[htb]
\centering
\small
\setlength{\tabcolsep}{6pt}
\renewcommand{\arraystretch}{1.25}
\caption{Event with the Central-resonant Degeneracy Reported Since 2021}
\begin{tabular}{l | c c c c c}
\hline
\hline
Event (Reference) 
& Classification
& $\Delta \log \rho$
& $\Delta \log q$
& $\Delta \chi^2 $  
& Datasets besides KMTNet \\
\hline
KMT-2021-BLG-0171 \citep{KB210171}
& Type\,II
& 0.039
& $-$0.334
& 6.4 
& LCOGT, Possum\\
KMT-2021-BLG-0192 \citep{KMT2021_mass3}
& Type\,II
& 0.073
& $-$0.228
& 23.4 
& --- \\
KMT-2021-BLG-1253 \citep{KMT2021_mass1}
& Type\,I
& $-$0.218
& $-$0.039
& $-$1.6
& --- \\
KMT-2021-BLG-1391 \citep{KMT2021_mass1}
& Type\,I
& $-$0.304
& $-$0.109
& $-$5.1 
& --- \\
KMT-2021-BLG-1689 \citep{KB210171}
& Type\,I
& $-$0.326
& $-$0.111
& 2.5
& Auckland Observatory, MOA \\
KMT-2022-BLG-0440 \citep{KB220440}
& Type\,II
& $>$0.196 ($1\sigma$)
& $-$0.684
& 91.1
& LCOGT, CHI-18 \\
KMT-2023-BLG-1431 \citep{KB231431}
& Type\,II
& 0.173
& $-$0.335
& 32.9
& LCOGT\\
KMT-2025-BLG-0811 (this work)
& Type\,I
& $-$0.242
& 0.058
& $-$7.8
& LCOGT\\
KMT-2025-BLG-1616 \citep{KB251616}
& Type\,I
& $-$0.379
& 0.013
& $-$329.0
& DECam \\

\hline
\hline
\end{tabular}
\begin{flushleft}
\textbf{Notes.} $\Delta \log \rho = \log \rho_{\rm resonant} - \log \rho_{\rm central}$; $\Delta \log q = \log q_{\rm resonant} - \log q_{\rm central}$; $\Delta\chi^2 = \chi^2_{\rm resonant} - \chi^2_{\rm central}$.
\end{flushleft}
\label{tab:central_resonant_summary}
\end{table*}

\begin{figure}[htb] 
    \centering
    \includegraphics[width=0.45\textwidth]{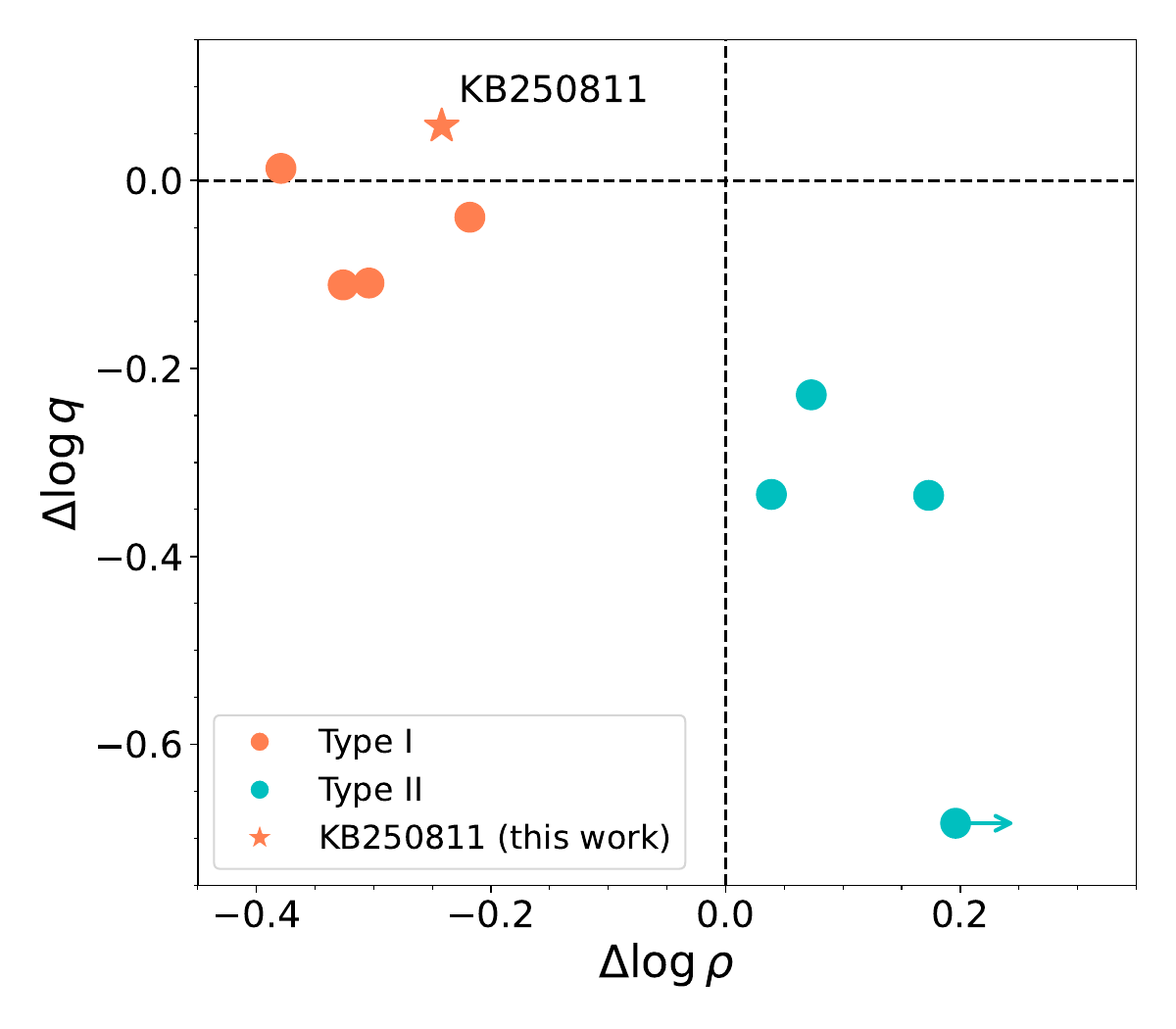}
    \caption{Distribution of the ``central-resonant'' degeneracies in the $(\Delta \log \rho, \Delta \log q)$ plane. Type\,I (orange) and Type\,II (cyan) events occupy distinct regions. The star marks \eventa\ (KB250811). The arrow indicates that the value of $\Delta \log \rho$ represents a $1\sigma$ lower limit, because for the ``central'' solutions of KMT-2022-BLG-0440 only an upper limit on $\rho$ is measured. Dashed lines denote $\Delta \log \rho = 0$ and $\Delta \log q = 0$.}
    \label{fig:dlogq_vs_dlogrho}
\end{figure}

\subsection{Two Types of Degeneracy}

In this work, we have reported the discovery and analysis of two low-$q$ microlensing planets, \eventa\ and \eventb, identified from two HM events. For \eventa, the planetary signal appears as a bump-type anomaly near the 1L1S peak. The 2L1S modeling reveals a ``central-resonant'' degeneracy which, when combined with the well-known ``close-wide'' degeneracy, results in four degenerate solutions. The ``central-resonant'' degeneracy was first systematically characterized in the 2021 observing season, during which five such events were identified. Since then, four additional events, including \eventa\ reported in this paper, have been found. For \eventa, the ``central-resonant'' solutions yield similar mass ratios of $q \sim 4.5 \times 10^{-5}$ but substantially different normalized source sizes $\rho$, and hence different angular Einstein radii and lens-source relative proper motions. To further investigate the ``central-resonant'' degeneracy, we review the nine such events detected since the 2021 season\footnote{We focus only on the ``central-resonant'' cases since the 2021 season because planetary events from this period have been investigated in detail, including careful exploration of local minima using methods such as the grid search approach described in \cite{KB220440}. As far as we are aware, the only previously reported case exhibiting this degeneracy is MOA-2011-BLG-262 \citep{MB11262}, which corresponds to our Type~I classification. We believe that the analyses of earlier events probably missed this degeneracy, as no such cases were reported in the 2016--2019 KMTNet survey.}.

Table~\ref{tab:central_resonant_summary} lists the event names, $\Delta \log \rho = \log \rho_{\rm resonant} - \log \rho_{\rm central}$, $\Delta \log q = \log q_{\rm resonant} - \log q_{\rm central}$, $\Delta\chi^2 = \chi^2_{\rm resonant} - \chi^2_{\rm central}$, and the datasets that cover the anomaly in addition to the KMTNet data, ordered by the event discovery date. Figure~\ref{fig:dlogq_vs_dlogrho} displays the distribution of $\Delta \log \rho$ versus $\Delta \log q$. Based on these quantities, two distinct types can be identified.

For the first type of this degeneracy, which we label as Type~I, the solutions have similar mass ratios, with $|\Delta \log q| \lesssim 0.1$, and the ``resonant'' solutions have significantly smaller normalized source sizes, with $\Delta \log \rho < -0.2$. This category contains five events, including \eventa. Based on their caustic geometries and source trajectories, the Type~I degeneracy arises when a relatively larger source crosses a central caustic, whereas a smaller source crosses a resonant caustic. In the light curves, the ``central'' solutions exhibit a single bump, while the ``resonant'' solutions display a double-peaked, U-shaped feature.

Four events fall into the second type of this degeneracy, which we label Type~II. The ``resonant'' solutions have relatively lower mass ratios, with $\Delta \log q < -0.2$, and larger normalized source sizes, $\Delta \log \rho > 0$. Based on their caustic geometries and source trajectories, the ``central'' solutions arise from a source cusp approach to the central caustic, whereas the ``resonant'' solutions, similar to Type~I, involve a source crossing. Because the ``central'' solutions do not involve a direct source crossing, a higher mass ratio is required to produce an anomaly with a similar amplitude. Moreover, because the extended area of the magnification spike is wider than the width of the resonant caustic, a smaller source size is needed to reproduce an anomaly with a comparable duration. In the light curves, both solutions produce a single bump. However, in the absence of a caustic crossing, the entry and exit of the bump for the ``central'' solutions are smooth, while the ``resonant'' solutions show dips there due to transverse cusp approaches.

\subsection{Resolution of the Degeneracy}

If we adopt a criterion of $\Delta\chi^2 > 10$ for resolving the ``central-resonant'' degeneracy, then among the four Type~II events, three have the degeneracy resolved by the observed data. Two of them, KMT-2021-BLG-0192 and KMT-2022-BLG-0440, can be resolved by the KMTNet survey data alone, with $\Gamma \geq 3~\text{hr}^{-1}$. For the third event, KMT-2023-BLG-1431, the KMTNet survey data with $\Gamma = 1 ~\text{hr}^{-1}$ are insufficient to break the degeneracy, and the dense follow-up observations from LCOGT and KMTNet, with a combined cadence of $\Gamma > 10~\text{hr}^{-1}$, resolved it. For the remaining event, KMT-2021-BLG-0171, the KMTNet and LCOGT data only covered about half of the anomaly. The Possum data covered the other half, but their photometric precision is insufficient to break the degeneracy. 

For the five Type~I events, only one has the ``central-resonant'' degeneracy resolved. Two of them, KMT-2021-BLG-1253 and KMT-2021-BLG-1391, have KMTNet survey data with $\Gamma \geq 3~\text{hr}^{-1}$. KMT-2021-BLG-1689 has $\Gamma \sim 30~\text{hr}^{-1}$ data from Auckland Observatory and $\Gamma \sim 1~\text{hr}^{-1}$ survey data from MOA, but the signal-to-noise ratio of each Auckland Observatory exposure is only about one quarter that of a typical KMTNet exposure.
\eventa\ is the most surprising case. It has a combined cadence of $\Gamma > 30~\text{hr}^{-1}$ from LCOGT and KMTNet, covering the entire anomaly, yet the degeneracy remains unresolved, with $\Delta\chi^2 = -7.8$. The only resolved case is KMT-2025-BLG-1616: the KMTNet survey data with $\Gamma = 4~\text{hr}^{-1}$ are insufficient to break the degeneracy, whereas $\Gamma = 24~\text{hr}^{-1}$ observations from DECam on the 4 m Blanco telescope (Yang et al. 2026, in prep.) successfully resolve it.

Therefore, compared to Type~II, Type~I of the ``central-resonant'' degeneracy is more difficult to resolve. This is reflected in their light curves. For Type~II, the differences between the two solutions occur mainly at the entry and exit of the bump, and they are more pronounced and last longer. In contrast, for Type~I, the differences are concentrated near the peak of the bump. The ``resonant'' solutions show an additional dip, but the overall deviations are weak. For \eventa, the differences are at most 0.03~mag and last for only about 20~min. 

Our review of the ``central-resonant'' degeneracy has two implications for upcoming microlensing surveys conducted by the Nancy Grace Roman Space Telescope (Roman) \citep{WFIRST,MatthewWFIRSTI} and the Earth~2.0 (ET) Microlensing Telescope \citep{CMST,ET}. First, the planned cadence for Roman is $\sim 5~\text{hr}^{-1}$ and for ET is $6~\text{hr}^{-1}$, which may not be sufficient to break the ``central-resonant'' degeneracy, especially for Type~I events. On the positive side, ET will have $\sim 22~\text{hr}^{-1}$ contemporaneous $I$-band observations from KMTNet, and the high-resolution imaging from Roman may enable measurements of the lens-source relative proper motion, which could resolve the Type~I degeneracy. In addition, satellite microlensing parallax \citep{1966MNRAS.134..315R,1994ApJ...421L..75G} from ET and KMTNet contemporaneous observations may provide further constraints on the degeneracy, although whether this will be sufficient to break the degeneracy remains uncertain.

Second, our summary of $\Delta \log q$ and $\Delta \log \rho$ in Table~\ref{tab:central_resonant_summary} and Figure~\ref{fig:dlogq_vs_dlogrho}, together with the classification into two types, provides guidance for searching for the alternative solution once one solution has been identified. This is particularly useful in the common situation in which the ``central'' solution is identified first (see Section~\ref{dis:phase}). Possibly this classification system could also be useful to find cases of this degeneracy in previous planetary events, especially for the KMTNet 2016--2019 sample. In addition, our results may offer guidance for theoretical interpretations of the ``central-resonant'' degeneracy from the perspective of the lens equation.

\subsection{Phase-space Factor}\label{dis:phase}

For degenerate 2L1S solutions, \cite{KB210171} suggested that the relative likelihood of each solution can be evaluated by applying phase-space factors \citep{OB110173} in $(\log s, \log q, \alpha)$. For the Type~I events, the ratio of the phase-space factors between the ``central'' and ``resonant'' solutions is $\sim 10$ for both \eventa\ and KMT-2021-BLG-1689, while the corresponding ratios were not reported in the papers for the other Type~I events. For the Type~II events, the ratio is higher, $\sim 60$, for both KMT-2021-BLG-0171 and KMT-2022-BLG-0440. Therefore, the ``central'' solutions are more likely to be identified in the grid search.

Based on the phase-space factors, the ``central'' solutions are expected to be 10--60 times more common than the ``resonant'' solutions. For the three Type~II events for which the degeneracy is resolved, the preferred solutions are all ``central''. However, in the only Type~I case for which the degeneracy has been broken, the correct solution is ``resonant''. The observed overall ratio of ``central'' to ``resonant'' solutions is 3:1, which is an order of magnitude smaller than the prediction of the phase-space factors. Although our current investigation is limited by the small sample size, we suggest that the phase-space factors should be carefully examined before being adopted in the construction of statistical samples.

\acknowledgments
Yuchen.T., W.Zang., H.Y., Z.L., J.Z., H.L., S.M., Q.Q. and W.Zhu acknowledge support by the National Natural Science Foundation of China (Grant No. 12133005). This research has made use of the KMTNet system operated by the Korea Astronomy and Space Science Institute (KASI) at three host sites of CTIO in Chile, SAAO in South Africa, and SSO in Australia. Data transfer from the host site to KASI was supported by the Korea Research Environment Open NETwork (KREONET).  This research was supported by KASI under the R\&D program (project No. 2025-1-830-05) supervised by the Ministry of Science and ICT. This work makes use of observations from the Las Cumbres Observatory global telescope network. The OGLE project has received funding from the Polish National Science Centre grant OPUS 2024/55/B/ST9/00447 awarded to A.U. H.Y. acknowledges support by the China Postdoctoral Science Foundation (No. 2024M762938). This work is part of the ET space mission which is funded by the China's Space Origins Exploration Program. RP acknowledges support by the Polish National Agency for Academic Exchange grant ``Polish Returns 2019''. Work by J.C.Y. acknowledges support from N.S.F Grant No. AST-2108414. Work by C.H. was supported by the grants of National Research Foundation of Korea (2019R1A2C2085965 and 2020R1A4A2002885). Y.S. acknowledges support from BSF Grant No. 2020740. W.Zhu acknowledges the science research grants from the China Manned Space Project with No.\ CMS-CSST-2021-A11. This work makes partial use of observations carried out at the Pico dos Dias Observatory (OPD), which is operated by the National Laboratory for Astrophysics (LNA).

\software{pySIS \citep{pysis,Yang_TLC,Yang_TLC2}, numpy \citep{numpy}, emcee \citep{emcee2,emcee}, Matplotlib \citep{Matplotlib}, SciPy \citep{scipy}, galpy \citep{bovy2015galpy}}

\bibliography{Zang.bib}

@article{WFIRST,
  title={Wide-field infrarred survey telescope-astrophysics focused telescope assets WFIRST-AFTA 2015 report},
  author={Spergel, Davidetal and Gehrels, Neil and Baltay, Charles and Bennett, David and Breckinridge, James and Donahue, Megan and Dressler, Alan and Gaudi, B Scott and Greene, Tom and Guyon, Olivier and others},
  journal={arXiv preprint arXiv:1503.03757},
  year={2015}
}

@ARTICLE{KB251616,
       author = {{Yang}, Hongjing and {Zang}, Weicheng and {Ryu}, Yoon-Hyun and {Sumi}, Takahiro and {Zhang}, Jiyuan and {Li}, Hongyu and {Han}, Cheongho and {Tang}, Yuchen and {Qian}, Qiyue and {Li}, Zhixing and {Shang}, Yuxin and {Shan}, Xikai and {Mao}, Shude and {Damke}, Guillermo and {Zenteno}, Alfredo and {Heathcote}, Steve and {Boutsia}, Konstantina and {Mr{\'o}z}, Przemek and {Zhao}, Xiurui and {Penny}, Matthew and {Terry}, Sean and {Tamburo}, Patrick and {Cunningham}, Timothy and {Ye}, Quanzhi and {Peng}, Eric W. and {Street}, Rachel and {Kruszy{\'n}ska}, Katarzyna and {Bachelet}, Etienne and {Tsapras}, Yiannis and {Hundertmark}, Markus and {Dreams Collaboration} and {Albrow}, Michael D. and {Chung}, Sun-Ju and {Gould}, Andrew and {Hwang}, Kyu-Ha and {Jung}, Youn Kil and {Shin}, In-Gu and {Shvartzvald}, Yossi and {Yee}, Jennifer C. and {Kim}, Dong-Jin and {Lee}, Chung-Uk and {Park}, Byeong-Gon and {Kmtnet Collaboration} and {Bennett}, David P. and {Bond}, Ian A. and {Hamada}, Ryusei and {Hirao}, Yuki and {Idei}, Asahi and {Makida}, Shuma and {Miyazaki}, Shota and {Nagai}, Tutumi and {Nagano}, Togo and {Nakayama}, Seiya and {Nishio}, Mayu and {Nunota}, Kansuke and {Ogawa}, Ryo and {Oishi}, Ryunosuke and {Okumoto}, Yui and {Rattenbury}, Nicholas J. and {Satoh}, Yuki K. and {Suzuki}, Daisuke and {Tamura}, Motohide and {Tamaoki}, Takuto and {Yama}, Hibiki and {Prime Collaboration}},
        title = "{KMT-2025-BLG-1616Lb: First Microlensing Bound Planet From DREAMS}",
      journal = {\aj},
         year = 2026,
        month = mar,
       volume = {171},
       number = {3},
          eid = {151},
        pages = {151},
          doi = {10.3847/1538-3881/ae36ac},
       adsurl = {https://ui.adsabs.harvard.edu/abs/2026AJ....171..151Y}
}

@article{Einstein1936,
author = {Albert Einstein },
title = {Lens-Like Action of a Star by the Deviation of Light in the Gravitational Field},
journal = {Science},
volume = {84},
number = {2188},
pages = {506-507},
year = {1936},
doi = {10.1126/science.84.2188.506},
URL = {https://www.science.org/doi/abs/10.1126/science.84.2188.506},
eprint = {https://www.science.org/doi/pdf/10.1126/science.84.2188.506}}

@ARTICLE{Three_planet_candidates,
       author = {{Zhang}, Jiyuan and {Zang}, Weicheng and {Ryu}, Yoon-Hyun and {Sumi}, Takahiro and {Udalski}, Andrzej and {Mao}, Shude and {Albrow}, Michael D. and {Chung}, Sun-Ju and {Gould}, Andrew and {Han}, Cheongho and {Hwang}, Kyu-Ha and {Jung}, Youn Kil and {Shin}, In-Gu and {Shvartzvald}, Yossi and {Yee}, Jennifer C. and {Yang}, Hongjing and {Cha}, Sang-Mok and {Kim}, Dong-Jin and {Kim}, Seung-Lee and {Lee}, Chung-Uk and {Lee}, Dong-Joo and {Lee}, Yongseok and {Park}, Byeong-Gon and {Pogge}, Richard W. and {Tang}, Yunyi and {de Almeida}, Leandro and {Maoz}, Dan and {Qian}, Qiyue and {Zhu}, Wei and {Abe}, Fumio and {Bando}, Ken and {Bennett}, David P. and {Bhattacharya}, Aparna and {Bond}, Ian A. and {Fukui}, Akihiko and {Hamada}, Ryusei and {Hamada}, Shunya and {Hamasaki}, Naoto and {Hirao}, Yuki and {Silva}, Stela Ishitani and {Koshimoto}, Naoki and {Matsubara}, Yutaka and {Miyazaki}, Shota and {Muraki}, Yasushi and {Nagai}, Tutumi and {Nunota}, Kansuke and {Olmschenk}, Greg and {Ranc}, Cl{\'e}ment and {Rattenbury}, Nicholas J. and {Satoh}, Yuki and {Suzuki}, Daisuke and {Terry}, Sean K. and {Tristram}, Paul J. and {Vandorou}, Aikaterini and {Yama}, Hibiki and {Mr{\'o}z}, Przemek and {Szyma{\'n}ski}, Micha{\l} K. and {Skowron}, Jan and {Poleski}, Radoslaw and {Soszy{\'n}ski}, Igor and {Pietrukowicz}, Pawe{\l} and {Koz{\l}owski}, Szymon and {Rybicki}, Krzysztof A. and {Iwanek}, Patryk and {Ulaczyk}, Krzysztof and {Wrona}, Marcin and {Gromadzki}, Mariusz and {Mr{\'o}z}, Mateusz J.},
        title = "{A comprehensive analysis of three microlensing planet candidates with the planet/binary degeneracy}",
      journal = {\mnras},
         year = 2026,
        month = feb,
       volume = {545},
       number = {4},
          eid = {staf1893},
        pages = {staf1893},
          doi = {10.1093/mnras/staf1893},
archivePrefix = {arXiv},
       eprint = {2509.18345},
 primaryClass = {astro-ph.EP},
       adsurl = {https://ui.adsabs.harvard.edu/abs/2026MNRAS.545S1893Z}
}

@ARTICLE{bovy2015galpy,
       author = {{Bovy}, Jo},
        title = "{galpy: A python Library for Galactic Dynamics}",
      journal = {\apjs},
         year = 2015,
        month = feb,
       volume = {216},
       number = {2},
          eid = {29},
        pages = {29},
          doi = {10.1088/0067-0049/216/2/29},
archivePrefix = {arXiv},
       eprint = {1412.3451},
 primaryClass = {astro-ph.GA},
       adsurl = {https://ui.adsabs.harvard.edu/abs/2015ApJS..216...29B}
}

@ARTICLE{PRIME,
       author = {{Sumi}, Takahiro and {Buckley}, David A.~H. and {Kutyrev}, Alexander S. and {Tamura}, Motohide and {Bennett}, David P. and {Bond}, Ian A. and {Cataldo}, Giuseppe and {Durbak}, Joseph M. and {Cenko}, S. Bradley and {Fixsen}, Dale and {Guiffreda}, Orion and {Hamada}, Ryusei and {Hirao}, Yuki and {Idei}, Asahi and {Kelly}, Dan and {Loose}, Markus and {Lotkin}, Gennadiy N. and {Lyness}, Eric I. and {Maher}, Stephen and {Makida}, Shuma and {Matsunaga}, Noriyuki and {Miyazaki}, Shota and {Mosby}, Gregory and {Moseley}, Samuel H. and {Nagai}, Tutumi and {Nagano}, Togo and {Nakayama}, Seiya and {Nishio}, Mayu and {Nunota}, Kansuke and {Ogawa}, Ryo and {Oishi}, Ryunosuke and {Okumoto}, Yui and {Rattenbury}, Nicholas J. and {Satoh}, Yuki K. and {Sharp}, Elmer H. and {Suzuki}, Daisuke and {Tamaoki}, Takuto and {Troja}, Eleonora and {White}, Sarah V. and {Yama}, Hibiki and {Prime Collaboration}},
        title = "{The PRime-focus Infrared Microlensing Experiment (PRIME): First Results}",
      journal = {\aj},
         year = 2025,
        month = dec,
       volume = {170},
       number = {6},
          eid = {338},
        pages = {338},
          doi = {10.3847/1538-3881/ae14f5},
archivePrefix = {arXiv},
       eprint = {2508.14474},
 primaryClass = {astro-ph.EP},
       adsurl = {https://ui.adsabs.harvard.edu/abs/2025AJ....170..338S}
}

@ARTICLE{NASAExo,
       author = {{Christiansen}, Jessie L. and {McElroy}, Douglas L. and {Harbut}, Marcy and {Ciardi}, David R. and {Crane}, Megan and {Good}, John and {Hardegree-Ullman}, Kevin K. and {Kesseli}, Aurora Y. and {Lund}, Michael B. and {Lynn}, Meca and {Muthiar}, Ananda and {Nilsson}, Ricky and {Oluyide}, Toba and {Papin}, Michael and {Rivera}, Amalia and {Swain}, Melanie and {Susemiehl}, Nicholas D. and {Tam}, Raymond and {van Eyken}, Julian and {Beichman}, Charles},
        title = "{The NASA Exoplanet Archive and Exoplanet Follow-up Observing Program: Data, Tools, and Usage}",
      journal = {PSJ},
         year = 2025,
        month = aug,
       volume = {6},
       number = {8},
          eid = {186},
        pages = {186},
          doi = {10.3847/PSJ/ade3c2},
archivePrefix = {arXiv},
       eprint = {2506.03299},
 primaryClass = {astro-ph.EP},
       adsurl = {https://ui.adsabs.harvard.edu/abs/2025PSJ.....6..186C}
}

@ARTICLE{OB160007,
       author = {{Zang}, Weicheng and {Jung}, Youn Kil and {Yee}, Jennifer C. and {Hwang}, Kyu-Ha and {Yang}, Hongjing and {Udalski}, Andrzej and {Sumi}, Takahiro and {Gould}, Andrew and {Mao}, Shude and {Albrow}, Michael D. and {Chung}, Sun-Ju and {Han}, Cheongho and {Ryu}, Yoon-Hyun and {Shin}, In-Gu and {Shvartzvald}, Yossi and {Cha}, Sang-Mok and {Kim}, Dong-Jin and {Kim}, Hyoun-Woo and {Kim}, Seung-Lee and {Lee}, Chung-Uk and {Lee}, Dong-Joo and {Lee}, Yongseok and {Park}, Byeong-Gon and {Pogge}, Richard W. and {Zhang}, Xiangyu and {Kuang}, Renkun and {Wang}, Hanyue and {Zhang}, Jiyuan and {Hu}, Zhecheng and {Zhu}, Wei and {Mr{\'o}z}, Przemek and {Skowron}, Jan and {Poleski}, Rados{\l}aw and {Szyma{\'n}ski}, Micha{\l} K. and {Soszy{\'n}ski}, Igor and {Pietrukowicz}, Pawe{\l} and {Koz{\l}owski}, Szymon and {Ulaczyk}, Krzysztof and {Rybicki}, Krzysztof A. and {Iwanek}, Patryk and {Wrona}, Marcin and {Gromadzki}, Mariusz and {Abe}, Fumio and {Barry}, Richard and {Bennett}, David P. and {Bhattacharya}, Aparna and {Bond}, Ian A. and {Fujii}, Hirosane and {Fukui}, Akihiko and {Hamada}, Ryusei and {Hirao}, Yuki and {Silva}, Stela Ishitani and {Itow}, Yoshitaka and {Kirikawa}, Rintaro and {Koshimoto}, Naoki and {Matsubara}, Yutaka and {Miyazaki}, Shota and {Muraki}, Yasushi and {Olmschenk}, Greg and {Ranc}, Cl{\'e}ment and {Rattenbury}, Nicholas J. and {Satoh}, Yuki and {Suzuki}, Daisuke and {Tomoyoshi}, Mio and {Tristram}, Paul J. and {Vandorou}, Aikaterini and {Yama}, Hibiki and {Yamashita}, Kansuke},
        title = "{Microlensing events indicate that super-Earth exoplanets are common in Jupiter-like orbits}",
      journal = {Science},
         year = 2025,
        month = apr,
       volume = {388},
       number = {6745},
        pages = {400-404},
          doi = {10.1126/science.adn6088},
archivePrefix = {arXiv},
       eprint = {2504.20158},
 primaryClass = {astro-ph.EP},
       adsurl = {https://ui.adsabs.harvard.edu/abs/2025Sci...388..400Z}
}

@ARTICLE{MASADA,
       author = {{Gould}, Andrew},
        title = "{MASADA: From Microlensing Planet Mass-Ratio Function to Planet Mass Function}",
      journal = {arXiv e-prints},
         year = 2022,
        month = sep,
          eid = {arXiv:2209.12501},
        pages = {arXiv:2209.12501},
archivePrefix = {arXiv},
       eprint = {2209.12501},
 primaryClass = {astro-ph.EP},
       adsurl = {https://ui.adsabs.harvard.edu/abs/2022arXiv220912501G}
}

@ARTICLE{Yang2021_GalacticModel,
       author = {{Yang}, Hongjing and {Mao}, Shude and {Zang}, Weicheng and {Zhang}, Xiangyu},
        title = "{Microlensing predictions: impact of Galactic disc dynamical models}",
      journal = {\mnras},
         year = 2021,
        month = apr,
       volume = {502},
       number = {4},
        pages = {5631-5642},
          doi = {10.1093/mnras/stab441},
archivePrefix = {arXiv},
       eprint = {2010.16146},
 primaryClass = {astro-ph.GA},
       adsurl = {https://ui.adsabs.harvard.edu/abs/2021MNRAS.502.5631Y}
}

@Article{Matplotlib,
  Author    = {Hunter, J. D.},
  Title     = {Matplotlib: A 2D graphics environment},
  Journal   = {Computing in Science \& Engineering},
  Volume    = {9},
  Number    = {3},
  Pages     = {90--95},
  publisher = {IEEE COMPUTER SOC},
  doi       = {10.1109/MCSE.2007.55},
  year      = 2007
}

@Article{numpy,
 title         = {Array programming with {NumPy}},
 author        = {Charles R. Harris and K. Jarrod Millman and St{\'{e}}fan J.
                 van der Walt and Ralf Gommers and Pauli Virtanen and David
                 Cournapeau and Eric Wieser and Julian Taylor and Sebastian
                 Berg and Nathaniel J. Smith and Robert Kern and Matti Picus
                 and Stephan Hoyer and Marten H. van Kerkwijk and Matthew
                 Brett and Allan Haldane and Jaime Fern{\'{a}}ndez del
                 R{\'{i}}o and Mark Wiebe and Pearu Peterson and Pierre
                 G{\'{e}}rard-Marchant and Kevin Sheppard and Tyler Reddy and
                 Warren Weckesser and Hameer Abbasi and Christoph Gohlke and
                 Travis E. Oliphant},
 year          = {2020},
 month         = sep,
 journal       = {Nature},
 volume        = {585},
 number        = {7825},
 pages         = {357--362},
 doi           = {10.1038/s41586-020-2649-2},
 publisher     = {Springer Science and Business Media {LLC}},
 url           = {https://doi.org/10.1038/s41586-020-2649-2}
}

@ARTICLE{Gould2000,
   author = {{Gould}, A.},
    title = "{A Natural Formalism for Microlensing}",
  journal = {\apj},
   eprint = {astro-ph/0001421},
     year = 2000,
    month = oct,
   volume = 542,
    pages = {785-788},
      doi = {10.1086/317037},
   adsurl = {http://adsabs.harvard.edu/abs/2000ApJ...542..785G}
}

@ARTICLE{Nataf2013,
   author = {{Nataf}, D.~M. and {Gould}, A. and {Fouqu{\'e}}, P. and {Gonzalez}, O.~A. and 
	{Johnson}, J.~A. and {Skowron}, J. and {}, A. and {Szyma{\'n}ski}, M.~K. and 
	{Kubiak}, M. and {Pietrzy{\'n}ski}, G. and {Soszy{\'n}ski}, I. and 
	{Ulaczyk}, K. and {Wyrzykowski}, {\L}. and {Poleski}, R.},
    title = "{Reddening and Extinction toward the Galactic Bulge from OGLE-III: The Inner Milky Way's R$_{V}$ \~{} 2.5 Extinction Curve}",
  journal = {\apj},
archivePrefix = "arXiv",
   eprint = {1208.1263},
     year = 2013,
    month = jun,
   volume = 769,
      eid = {88},
    pages = {88},
      doi = {10.1088/0004-637X/769/2/88},
   adsurl = {http://adsabs.harvard.edu/abs/2013ApJ...769...88N}
}

@ARTICLE{1994ApJ...421L..75G,
   author = {{Gould}, A.},
    title = "{MACHO velocities from satellite-based parallaxes}",
  journal = {\apjl},
     year = 1994,
    month = feb,
   volume = 421,
    pages = {L75-L78},
      doi = {10.1086/187191},
   adsurl = {http://adsabs.harvard.edu/abs/1994ApJ...421L..75G}
}

@ARTICLE{Zhu2017spitzer,
   author = {{Zhu}, W. and {Udalski}, A. and {Calchi Novati}, S. and {Chung}, S.-J. and 
	{Jung}, Y.~K. and {Ryu}, Y.-H. and {}, I.-G. and {Gould}, A. and 
	{Lee}, C.-U. and {Albrow}, M.~D. and {Yee}, J.~C. and {Han}, C. and 
	{Hwang}, K.-H. and {Cha}, S.-M. and {Kim}, D.-J. and {Kim}, H.-W. and 
	{Kim}, S.-L. and {Kim}, Y.-H. and {Lee}, Y. and {Park}, B.-G. and 
	{Pogge}, R.~W. and {KMTNet Collaboration} and {Poleski}, R. and 
	{Mr{\'o}z}, P. and {Pietrukowicz}, P. and {Skowron}, J. and 
	{Szyma{\'n}ski}, M.~K. and {Koz{\l}owski}, S. and {Ulaczyk}, K. and 
	{Pawlak}, M. and {OGLE Collaboration} and {Beichman}, C. and 
	{Bryden}, G. and {Carey}, S. and {Fausnaugh}, M. and {Gaudi}, B.~S. and 
	{Henderson}, C.~B. and {Shvartzvald}, Y. and {Wibking}, B. and 
	{Spitzer Team}},
    title = "{Toward a Galactic Distribution of Planets. I. Methodology and Planet Sensitivities of the 2015 High-cadence Spitzer Microlens Sample}",
  journal = {\aj},
archivePrefix = "arXiv",
   eprint = {1701.05191},
 primaryClass = "astro-ph.EP",
     year = 2017,
    month = nov,
   volume = 154,
      eid = {210},
    pages = {210},
      doi = {10.3847/1538-3881/aa8ef1},
   adsurl = {http://adsabs.harvard.edu/abs/2017AJ....154..210Z}
}

@ARTICLE{Alard1998,
   author = {{Alard}, C. and {Lupton}, R.~H.},
    title = "{A Method for Optimal Image Subtraction}",
  journal = {\apj},
   eprint = {astro-ph/9712287},
     year = 1998,
    month = aug,
   volume = 503,
    pages = {325-331},
      doi = {10.1086/305984},
   adsurl = {http://adsabs.harvard.edu/abs/1998ApJ...503..325A}
}

@ARTICLE{Udalski2003,
   author = {{Udalski}, A.},
    title = "{The Optical Gravitational Lensing Experiment. Real Time Data Analysis Systems in the OGLE-III Survey}",
  journal = {\actaa},
   eprint = {astro-ph/0401123},
     year = 2003,
    month = dec,
   volume = 53,
    pages = {291-305},
   adsurl = {http://adsabs.harvard.edu/abs/2003AcA....53..291U}
}

@ARTICLE{KB210912,
       author = {{Han}, Cheongho and {Bond}, Ian A. and {Yee}, Jennifer C. and {Zang}, Weicheng and {Albrow}, Michael D. and {Chung}, Sun-Ju and {Gould}, Andrew P. and {Hwang}, Kyu-Ha and {Jung}, Youn Kil and {Kim}, Doeon and {Lee}, Chung-Uk and {Ryu}, Yoon-Hyun and {Shin}, In-Gu and {Shvartzvald}, Yossi and {Cha}, Sang-Mok and {Kim}, Dong-Jin and {Kim}, Seung-Lee and {Lee}, Dong-Joo and {Lee}, Yongseok and {Park}, Byeong-Gon and {Pogge}, Richard W. and {Abe}, Fumio and {Barry}, Richard and {Bennett}, David P. and {Bhattacharya}, Aparna and {Hirao}, Yuki and {Fujii}, Hirosane and {Fukui}, Akihiko and {Itow}, Yoshitaka and {Kirikawa}, Rintaro and {Kondo}, Iona and {Koshimoto}, Naoki and {Matsubara}, Yutaka and {Matsumoto}, Sho and {Muraki}, Yasushi and {Miyazaki}, Shota and {Ranc}, Cl{\'e}ment and {Okamura}, Arisa and {Rattenbury}, Nicholas J. and {Satoh}, Yuki and {Sumi}, Takahiro and {Suzuki}, Daisuke and {Ishitani Silva}, Stela and {Toda}, Taiga and {Tristram}, Paul J. and {Yama}, Hibiki and {Yonehara}, Atsunori and {Cooper}, Tony and {Dimitrov}, Plamen and {Dong}, Subo and {Drummond}, John and {Green}, Jonathan and {Hennerley}, Steve and {Liu}, Zhuokai and {Mao}, Shude and {Maoz}, Dan and {Penny}, Matthew and {Yang}, Hongjing},
        title = "{KMT-2021-BLG-0912Lb: a microlensing super Earth around a K-type star}",
      journal = {\aap},
         year = 2022,
        month = feb,
       volume = {658},
          eid = {A94},
        pages = {A94},
          doi = {10.1051/0004-6361/202142495},
       adsurl = {https://ui.adsabs.harvard.edu/abs/2022A&A...658A..94H}
}

@ARTICLE{Paczynski1986,
   author = {{Paczy{\'n}ski}, B.},
    title = "{Gravitational microlensing by the galactic halo}",
  journal = {\apj},
     year = 1986,
    month = may,
   volume = 304,
    pages = {1-5},
      doi = {10.1086/164140},
   adsurl = {http://adsabs.harvard.edu/abs/1986ApJ...304....1P}
}

@ARTICLE{Adams2018,
   author = {{Adams}, A.~D. and {Boyajian}, T.~S. and {von Braun}, K.},
    title = "{Predicting stellar angular diameters from V, I$_{C}$, H and K photometry}",
  journal = {\mnras},
archivePrefix = "arXiv",
   eprint = {1709.03902},
 primaryClass = "astro-ph.SR",
     year = 2018,
    month = jan,
   volume = 473,
    pages = {3608-3614},
      doi = {10.1093/mnras/stx2367},
   adsurl = {http://adsabs.harvard.edu/abs/2018MNRAS.473.3608A}
}

@ARTICLE{ET,
       author = {{Ge}, Jian and {Zhang}, Hui and {Zang}, Weicheng and {Deng}, Hongping and {Mao}, Shude and {Xie}, Ji-Wei and {Liu}, Hui-Gen and {Zhou}, Ji-Lin and {Willis}, Kevin and {Huang}, Chelsea and {Howell}, Steve B. and {Feng}, Fabo and {Zhu}, Jiapeng and {Yao}, Xinyu and {Liu}, Beibei and {Aizawa}, Masataka and {Zhu}, Wei and {Li}, Ya-Ping and {Ma}, Bo and {Ye}, Quanzhi and {Yu}, Jie and {Xiang}, Maosheng and {Yu}, Cong and {Liu}, Shangfei and {Yang}, Ming and {Wang}, Mu-Tian and {Shi}, Xian and {Fang}, Tong and {Zong}, Weikai and {Liu}, Jinzhong and {Zhang}, Yu and {Zhang}, Liyun and {El-Badry}, Kareem and {Shen}, Rongfeng and {Tam}, Pak-Hin Thomas and {Hu}, Zhecheng and {Yang}, Yanlv and {Zou}, Yuan-Chuan and {Wu}, Jia-Li and {Lei}, Wei-Hua and {Wei}, Jun-Jie and {Wu}, Xue-Feng and {Sun}, Tian-Rui and {Wang}, Fa-Yin and {Zhang}, Bin-Bin and {Xu}, Dong and {Yang}, Yuan-Pei and {Li}, Wen-Xiong and {Xiang}, Dan-Feng and {Wang}, Xiaofeng and {Wang}, Tinggui and {Zhang}, Bing and {Jia}, Peng and {Yuan}, Haibo and {Zhang}, Jinghua and {Xuesong Wang}, Sharon and {Gan}, Tianjun and {Wang}, Wei and {Zhao}, Yinan and {Liu}, Yujuan and {Wei}, Chuanxin and {Kang}, Yanwu and {Yang}, Baoyu and {Qi}, Chao and {Liu}, Xiaohua and {Zhang}, Quan and {Zhu}, Yuji and {Zhou}, Dan and {Zhang}, Congcong and {Yu}, Yong and {Zhang}, Yongshuai and {Li}, Yan and {Tang}, Zhenghong and {Wang}, Chaoyan and {Wang}, Fengtao and {Li}, Wei and {Cheng}, Pengfei and {Shen}, Chao and {Li}, Baopeng and {Pan}, Yue and {Yang}, Sen and {Gao}, Wei and {Song}, Zongxi and {Wang}, Jian and {Zhang}, Hongfei and {Chen}, Cheng and {Wang}, Hui and {Zhang}, Jun and {Wang}, Zhiyue and {Zeng}, Feng and {Zheng}, Zhenhao and {Zhu}, Jie and {Guo}, Yingfan and {Zhang}, Yihao and {Li}, Yudong and {Wen}, Lin and {Feng}, Jie and {Chen}, Wen and {Chen}, Kun and {Han}, Xingbo and {Yang}, Yingquan and {Wang}, Haoyu and {Duan}, Xuliang and {Huang}, Jiangjiang and {Liang}, Hong and {Bi}, Shaolan and {Gai}, Ning and {Ge}, Zhishuai and {Guo}, Zhao and {Huang}, Yang and {Li}, Gang and {Li}, Haining and {Li}, Tanda and {Yuxi} and {Lu} and {Rix}, Hans-Walter and {Shi}, Jianrong and {Song}, Fen and {Tang}, Yanke and {Ting}, Yuan-Sen and {Wu}, Tao and {Wu}, Yaqian and {Yang}, Taozhi and {Yin}, Qing-Zhu and {Gould}, Andrew and {Lee}, Chung-Uk and {Dong}, Subo and {Yee}, Jennifer C. and {Shvartzvald}, Yossi and {Yang}, Hongjing and {Kuang}, Renkun and {Zhang}, Jiyuan and {Liao}, Shilong and {Qi}, Zhaoxiang and {Yang}, Jun and {Zhang}, Ruisheng and {Jiang}, Chen and {Ou}, Jian-Wen and {Li}, Yaguang and {Beck}, Paul and {Bedding}, Timothy R. and {Campante}, Tiago L. and {Chaplin}, William J. and {Christensen-Dalsgaard}, J{\o}rgen and {Garc{\'\i}a}, Rafael A. and {Gaulme}, Patrick and {Gizon}, Laurent and {Hekker}, Saskia and {Huber}, Daniel and {Khanna}, Shourya and {Li}, Yan and {Mathur}, Savita and {Miglio}, Andrea and {Mosser}, Beno{\^\i}t and {Ong}, J.~M. Joel and {Santos}, {\^A}ngela R.~G. and {Stello}, Dennis and {Bowman}, Dominic M. and {Lares-Martiz}, Mariel and {Murphy}, Simon and {Niu}, Jia-Shu and {Ma}, Xiao-Yu and {Moln{\'a}r}, L{\'a}szl{\'o} and {Fu}, Jian-Ning and {De Cat}, Peter and {Su}, Jie and {consortium}, the ET},
        title = "{ET White Paper: To Find the First Earth 2.0}",
      journal = {arXiv e-prints},
         year = 2022,
        month = jun,
          eid = {arXiv:2206.06693},
        pages = {arXiv:2206.06693},
          doi = {10.48550/arXiv.2206.06693},
archivePrefix = {arXiv},
       eprint = {2206.06693},
 primaryClass = {astro-ph.IM},
       adsurl = {https://ui.adsabs.harvard.edu/abs/2022arXiv220606693G}
}

@ARTICLE{emcee,
   author = {{Foreman-Mackey}, D. and {Hogg}, D.~W. and {Lang}, D. and {Goodman}, J.
	},
    title = "{emcee: The MCMC Hammer}",
  journal = {\pasp},
archivePrefix = "arXiv",
   eprint = {1202.3665},
 primaryClass = "astro-ph.IM",
     year = 2013,
    month = mar,
   volume = 125,
    pages = {306},
      doi = {10.1086/670067},
   adsurl = {http://adsabs.harvard.edu/abs/2013PASP..125..306F}
}

@ARTICLE{OGLEIII,
   author = {{Szyma{\'n}ski}, M.~K. and {Udalski}, A. and {Soszy{\'n}ski}, I. and 
	{Kubiak}, M. and {Pietrzy{\'n}ski}, G. and {Poleski}, R. and 
	{Wyrzykowski}, {\L}. and {Ulaczyk}, K.},
    title = "{The Optical Gravitational Lensing Experiment. OGLE-III Photometric Maps of the Galactic Bulge Fields}",
  journal = {\actaa},
archivePrefix = "arXiv",
   eprint = {1107.4008},
 primaryClass = "astro-ph.SR",
     year = 2011,
    month = jun,
   volume = 61,
    pages = {83-102},
   adsurl = {http://adsabs.harvard.edu/abs/2011AcA....61...83S}
}

@ARTICLE{Udalski1994,
   author = {{Udalski}, A. and {Szymanski}, M. and {Kaluzny}, J. and {Kubiak}, M. and 
	{Mateo}, M. and {Krzeminski}, W. and {Paczynski}, B.},
    title = "{The Optical Gravitational Lensing Experiment. The Early Warning System: Real Time Microlensing}",
  journal = {\actaa},
   eprint = {astro-ph/9408026},
     year = 1994,
    month = jul,
   volume = 44,
    pages = {227-234},
   adsurl = {http://adsabs.harvard.edu/abs/1994AcA....44..227U}
}

@ARTICLE{1966MNRAS.134..315R,
   author = {{Refsdal}, S.},
    title = "{On the possibility of determining the distances and masses of stars from the gravitational lens effect}",
  journal = {\mnras},
     year = 1966,
   volume = 134,
    pages = {315},
      doi = {10.1093/mnras/134.3.315},
   adsurl = {http://adsabs.harvard.edu/abs/1966MNRAS.134..315R}
}

@ARTICLE{KMT2016,
   author = {{Kim}, S.-L. and {Lee}, C.-U. and {Park}, B.-G. and {Kim}, D.-J. and 
	{Cha}, S.-M. and {Lee}, Y. and {Han}, C. and {Chun}, M.-Y. and 
	{Yuk}, I.},
    title = "{KMTNET: A Network of 1.6 m Wide-Field Optical Telescopes Installed at Three Southern Observatories}",
  journal = {Journal of Korean Astronomical Society},
     year = 2016,
    month = feb,
   volume = 49,
    pages = {37-44},
      doi = {10.5303/JKAS.2016.49.1.37},
   adsurl = {http://adsabs.harvard.edu/abs/2016JKAS...49...37K}
}

@ARTICLE{Shude1991,
   author = {{Mao}, S. and {Paczynski}, B.},
    title = "{Gravitational microlensing by double stars and planetary systems}",
  journal = {\apjl},
     year = 1991,
    month = jun,
   volume = 374,
    pages = {L37-L40},
      doi = {10.1086/186066},
   adsurl = {http://adsabs.harvard.edu/abs/1991ApJ...374L..37M}
}

@ARTICLE{Andy1992,
   author = {{Gould}, A. and {Loeb}, A.},
    title = "{Discovering planetary systems through gravitational microlenses}",
  journal = {\apj},
     year = 1992,
    month = sep,
   volume = 396,
    pages = {104-114},
      doi = {10.1086/171700},
   adsurl = {http://adsabs.harvard.edu/abs/1992ApJ...396..104G}
}

@ARTICLE{Mao2012,
   author = {{Mao}, S.},
    title = "{Astrophysical applications of gravitational microlensing}",
  journal = {Research in Astronomy and Astrophysics},
archivePrefix = "arXiv",
   eprint = {1207.3720},
     year = 2012,
    month = aug,
   volume = 12,
    pages = {947-972},
      doi = {10.1088/1674-4527/12/8/005},
   adsurl = {http://adsabs.harvard.edu/abs/2012RAA....12..947M}
}

@ARTICLE{Gaudi2012,
   author = {{Gaudi}, B.~S.},
    title = "{Microlensing Surveys for Exoplanets}",
  journal = {\araa},
     year = 2012,
    month = sep,
   volume = 50,
    pages = {411-453},
      doi = {10.1146/annurev-astro-081811-125518},
   adsurl = {http://adsabs.harvard.edu/abs/2012ARA%26A..50..411G}
}

@ARTICLE{OGLEIV,
   author = {{Udalski}, A. and {Szyma{\'n}ski}, M.~K. and {Szyma{\'n}ski}, G.
	},
    title = "{OGLE-IV: Fourth Phase of the Optical Gravitational Lensing Experiment}",
  journal = {\actaa},
archivePrefix = "arXiv",
   eprint = {1504.05966},
 primaryClass = "astro-ph.SR",
     year = 2015,
    month = mar,
   volume = 65,
    pages = {1-38},
   adsurl = {http://adsabs.harvard.edu/abs/2015AcA....65....1U}
}

@ARTICLE{Gaudi1998,
   author = {{Gaudi}, B.~S.},
    title = "{Distinguig Between Binary-Source and Planetary Microlensing Perturbations}",
  journal = {\apj},
     year = 1998,
    month = oct,
   volume = 506,
    pages = {533-539},
      doi = {10.1086/306256},
   adsurl = {http://adsabs.harvard.edu/abs/1998ApJ...506..533G}
}

@ARTICLE{pysis,
   author = {{Albrow}, M.~D. and {Horne}, K. and {Bramich}, D.~M. and {Fouqu{\'e}}, P. and 
	{Miller}, V.~R. and {Beaulieu}, J.-P. and {Coutures}, C. and 
	{Menzies}, J. and {Williams}, A. and {Batista}, V. and {Bennett}, D.~P. and 
	{Brillant}, S. and {Cassan}, A. and {Dieters}, S. and {Dominis Prester}, D. and 
	{Donatowicz}, J. and {Greenhill}, J. and {Kains}, N. and {Kane}, S.~R. and 
	{Kubas}, D. and {Marquette}, J.~B. and {Pollard}, K.~R. and 
	{Sahu}, K.~C. and {Tsapras}, Y. and {Wambsganss}, J. and {Zub}, M.
	},
    title = "{Difference imaging photometry of blended gravitational microlensing events with a numerical kernel}",
  journal = {\mnras},
archivePrefix = "arXiv",
   eprint = {0905.3003},
 primaryClass = "astro-ph.SR",
     year = 2009,
    month = aug,
   volume = 397,
    pages = {2099-2105},
      doi = {10.1111/j.1365-2966.2009.15098.x},
   adsurl = {http://adsabs.harvard.edu/abs/2009MNRAS.397.2099A}
}

@ARTICLE{Bozza2010,
   author = {{Bozza}, V.},
    title = "{Microlensing with an advanced contour integration algorithm: Green's theorem to third order, error control, optimal sampling and limb darkening}",
  journal = {\mnras},
archivePrefix = "arXiv",
   eprint = {1004.2796},
 primaryClass = "astro-ph.EP",
     year = 2010,
    month = nov,
   volume = 408,
    pages = {2188-2200},
      doi = {10.1111/j.1365-2966.2010.17265.x},
   adsurl = {http://adsabs.harvard.edu/abs/2010MNRAS.408.2188B}
}

@ARTICLE{MB12486,
   author = {{Hwang}, K.-H. and {Choi}, J.-Y. and {Bond}, I.~A. and {Sumi}, T. and 
	{Han}, C. and {Gaudi}, B.~S. and {Gould}, A. and {Bozza}, V. and 
	{Beaulieu}, J.-P. and {Tsapras}, Y. and {Abe}, F. and {Bennett}, D.~P. and 
	{Botzler}, C.~S. and {Chote}, P. and {Freeman}, M. and {Fukui}, A. and 
	{Fukunaga}, D. and {Harris}, P. and {Itow}, Y. and {Koshimoto}, N. and 
	{Ling}, C.~H. and {Masuda}, K. and {Matsubara}, Y. and {Muraki}, Y. and 
	{Namba}, S. and {Ohnishi}, K. and {Rattenbury}, N.~J. and {Saito}, T. and 
	{Sullivan}, D.~J. and {Sweatman}, W.~L. and {Suzuki}, D. and 
	{Tristram}, P.~J. and {Wada}, K. and {Yamai}, N. and {Yock}, P.~C.~M. and 
	{Yonehara}, A. and {MOA Collaboration} and {de Almeida}, L.~A. and 
	{DePoy}, D.~L. and {Dong}, S. and {Jablonski}, F. and {Jung}, Y.~K. and 
	{Kavka}, A. and {Lee}, C.-U. and {Park}, H. and {Pogge}, R.~W. and 
	{}, I.-G. and {Yee}, J.~C. and {{$\mu$}FUN Collaboration} and 
	{Albrow}, M.~D. and {Bachelet}, E. and {Batista}, V. and {Brillant}, S. and 
	{Caldwell}, J.~A.~R. and {Cassan}, A. and {Cole}, A. and {Corrales}, E. and 
	{Coutures}, C. and {Dieters}, S. and {Dominis Prester}, D. and 
	{Donatowicz}, J. and {Fouqu{\'e}}, P. and {Greenhill}, J. and 
	{J{\o}rgensen}, U.~G. and {Kane}, S.~R. and {Kubas}, D. and 
	{Marquette}, J.-B. and {Martin}, R. and {Meintjes}, P. and {Menzies}, J. and 
	{Pollard}, K.~R. and {Williams}, A. and {Wouters}, D. and {PLANET Collaboration} and 
	{Bramich}, D.~M. and {Dominik}, M. and {Horne}, K. and {Browne}, P. and 
	{Hundertmark}, M. and {Ipatov}, S. and {Kains}, N. and {Snodgrass}, C. and 
	{Steele}, I.~A. and {Street}, R.~A. and {RoboNet Collaboration}
	},
    title = "{Interpretation of a Short-term Anomaly in the Gravitational Microlensing Event MOA-2012-BLG-486}",
  journal = {\apj},
archivePrefix = "arXiv",
   eprint = {1308.5762},
 primaryClass = "astro-ph.SR",
     year = 2013,
    month = nov,
   volume = 778,
      eid = {55},
    pages = {55},
      doi = {10.1088/0004-637X/778/1/55},
   adsurl = {http://adsabs.harvard.edu/abs/2013ApJ...778...55H}
}

@ARTICLE{Yoo2004,
   author = {{Yoo}, J. and {DePoy}, D.~L. and {Gal-Yam}, A. and {Gaudi}, B.~S. and 
	{Gould}, A. and {Han}, C. and {Lipkin}, Y. and {Maoz}, D. and 
	{Ofek}, E.~O. and {Park}, B.-G. and {Pogge}, R.~W. and {Mu-Fun Collaboration} and 
	{Udalski}, A. and {Soszy{\'n}ski}, I. and {Wyrzykowski}, {\L}. and 
	{Kubiak}, M. and {Szyma{\'n}ski}, M. and {Pietrzy{\'n}ski}, G. and 
	{Szewczyk}, O. and {{\.Z}ebru{\'n}}, K. and {OGLE Collaboration}
	},
    title = "{OGLE-2003-BLG-262: Finite-Source Effects from a Point-Mass Lens}",
  journal = {\apj},
   eprint = {astro-ph/0309302},
     year = 2004,
    month = mar,
   volume = 603,
    pages = {139-151},
      doi = {10.1086/381241},
   adsurl = {http://adsabs.harvard.edu/abs/2004ApJ...603..139Y}
}

@ARTICLE{Houdashelt2000,
   author = {{Houdashelt}, M.~L. and {Bell}, R.~A. and {Sweigart}, A.~V.},
    title = "{Improved Color-Temperature Relations and Bolometric Corrections for Cool Stars}",
  journal = {\aj},
   eprint = {astro-ph/9911367},
     year = 2000,
    month = mar,
   volume = 119,
    pages = {1448-1469},
      doi = {10.1086/301243},
   adsurl = {http://adsabs.harvard.edu/abs/2000AJ....119.1448H}
}

@ARTICLE{Claret2011,
   author = {{Claret}, A. and {Bloemen}, S.},
    title = "{Gravity and limb-darkening coefficients for the Kepler, CoRoT, Spitzer, uvby, UBVRIJHK, and Sloan photometric systems}",
  journal = {\aap},
     year = 2011,
    month = may,
   volume = 529,
      eid = {A75},
    pages = {A75},
      doi = {10.1051/0004-6361/201116451},
   adsurl = {http://adsabs.harvard.edu/abs/2011A%26A...529A..75C}
}

@ARTICLE{Wozniak2000,
   author = {{Wozniak}, P.~R.},
    title = "{Difference Image Analysis of the OGLE-II Bulge Data. I. The Method}",
  journal = {\actaa},
   eprint = {astro-ph/0012143},
     year = 2000,
    month = dec,
   volume = 50,
    pages = {421-450},
   adsurl = {http://adsabs.harvard.edu/abs/2000AcA....50..421W}
}

@ARTICLE{snowline,
   author = {{Kennedy}, G.~M. and {Kenyon}, S.~J.},
    title = "{Planet Formation around Stars of Various Masses: The Snow Line and the Frequency of Giant Planets}",
  journal = {\apj},
archivePrefix = "arXiv",
   eprint = {0710.1065},
     year = 2008,
    month = jan,
   volume = 673,
      eid = {502-512},
    pages = {502-512},
      doi = {10.1086/524130},
   adsurl = {http://adsabs.harvard.edu/abs/2008ApJ...673..502K}
}

@ARTICLE{VBMicrolensing2025,
       author = {{Bozza}, V. and {Saggese}, V. and {Covone}, G. and {Rota}, P. and {Zhang}, J.},
        title = "{VBMicroLensing: Three algorithms for multiple lensing with contour integration}",
      journal = {\aap},
         year = 2025,
        month = feb,
       volume = {694},
          eid = {A219},
        pages = {A219},
          doi = {10.1051/0004-6361/202452648},
archivePrefix = {arXiv},
       eprint = {2410.13660},
 primaryClass = {astro-ph.IM},
       adsurl = {https://ui.adsabs.harvard.edu/abs/2025A&A...694A.219B}
}

@ARTICLE{KB231431,
       author = {{Bell}, Aislyn and {Zhang}, Jiyuan and {Zang}, Weicheng and {Jung}, Youn Kil and {Yee}, Jennifer C. and {Yang}, Hongjing and {Sumi}, Takahiro and {Udalski}, Andrzej and {Albrow}, Michael D. and {Chung}, Sun-Ju and {Gould}, Andrew and {Han}, Cheongho and {Hwang}, Kyu-Ha and {Ryu}, Yoon-Hyun and {Shin}, In-Gu and {Shvartzvald}, Yossi and {Cha}, Sang-Mok and {Kim}, Dong-Jin and {Kim}, Seung-Lee and {Lee}, Chung-Uk and {Lee}, Dong-Joo and {Lee}, Yongseok and {Park}, Byeong-Gon and {Pogge}, Richard W. and {KMTNet Collaboration} and {Tang}, Yunyi and {McCormick}, Jennie and {Dong}, Subo and {Liu}, Zhuokai and {de Almeida}, Leandro and {Mao}, Shude and {Maoz}, Dan and {Zhu}, Wei and {MAP and {\ensuremath{\mu}}FUN Follow-up Team} and {Abe}, Fumio and {Barry}, Richard and {Bennett}, David P. and {Bhattacharya}, Aparna and {Bond}, Ian A. and {Fujii}, Hirosane and {Fukui}, Akihiko and {Hamada}, Ryusei and {Hirao}, Yuki and {Silva}, Stela Ishitani and {Itow}, Yoshitaka and {Kirikawa}, Rintaro and {Kondo}, Iona and {Koshimoto}, Naoki and {Matsubara}, Yutaka and {Matsumoto}, Sho and {Miyazaki}, Shota and {Muraki}, Yasushi and {Okamura}, Arisa and {Lmschenk}, Greg and {Ranc}, Cl{\'e}ment and {Rattenbury}, Nicholas J. and {Satoh}, Yuki and {Suzuki}, Daisuke and {Toda}, Taiga and {Tomoyoshi}, Mio and {Tristram}, Paul J. and {Vandorou}, Aikaterini and {Yama}, Hibiki and {Yamashita}, Kansuke and {MOA Collaboration} and {Mr{\'o}z}, Przemek and {Skowron}, Jan and {Poleski}, Radoslaw and {Szyma{\'n}ski}, Micha{\l} K. and {Soszy{\'n}ski}, Igor and {Pietrukowicz}, Pawe{\l} and {Koz{\l}owski}, Szymon and {Ulaczyk}, Krzysztof and {Rybicki}, Krzysztof A. and {Iwanek}, Patryk and {Wrona}, Marcin and {Gromadzki}, Mariusz and {OGLE Collaboration}},
        title = "{KMT-2023-BLG-1431Lb: A New q < 10$^{‑4}$ Microlensing Planet from a Subtle Signature}",
      journal = {\pasp},
         year = 2024,
        month = may,
       volume = {136},
       number = {5},
          eid = {054402},
        pages = {054402},
          doi = {10.1088/1538-3873/ad48b8},
archivePrefix = {arXiv},
       eprint = {2311.13097},
 primaryClass = {astro-ph.EP},
       adsurl = {https://ui.adsabs.harvard.edu/abs/2024PASP..136e4402B}
}

@ARTICLE{KB220440,
       author = {{Zhang}, Jiyuan and {Zang}, Weicheng and {Jung}, Youn Kil and {Yang}, Hongjing and {Gould}, Andrew and {Sumi}, Takahiro and {Mao}, Shude and {Dong}, Subo and {Albrow}, Michael D. and {Chung}, Sun-Ju and {Han}, Cheongho and {Hwang}, Kyu-Ha and {Ryu}, Yoon-Hyun and {Shin}, In-Gu and {Shvartzvald}, Yossi and {Yee}, Jennifer C. and {Cha}, Sang-Mok and {Kim}, Dong-Jin and {Kim}, Hyoun-Woo and {Kim}, Seung-Lee and {Lee}, Chung-Uk and {Lee}, Dong-Joo and {Lee}, Yongseok and {Park}, Byeong-Gon and {Pogge}, Richard W. and {Qian}, Qiyue and {Liu}, Zhuokai and {Maoz}, Dan and {Penny}, Matthew T. and {Zhu}, Wei and {Abe}, Fumio and {Barry}, Richard and {Bennett}, David P. and {Bhattacharya}, Aparna and {Bond}, Ian A. and {Fujii}, Hirosane and {Fukui}, Akihiko and {Hamada}, Ryusei and {Hirao}, Yuki and {Silva}, Stela Ishitani and {Itow}, Yoshitaka and {Kirikawa}, Rintaro and {Kondo}, Iona and {Koshimoto}, Naoki and {Matsubara}, Yutaka and {Matsumoto}, Sho and {Miyazaki}, Shota and {Muraki}, Yasushi and {Okamura}, Arisa and {Olmschenk}, Greg and {Ranc}, Cl{\'e}ment and {Rattenbury}, Nicholas J. and {Satoh}, Yuki and {Suzuki}, Daisuke and {Toda}, Taiga and {Tomoyoshi}, Mio and {Tristram}, Paul J. and {Vandorou}, Aikaterini and {Yama}, Hibiki and {Yamashita}, Kansuke and {MOA Collaboration}},
        title = "{KMT-2022-BLG-0440Lb: A new q < 10$^{-4}$ microlensing planet with the central-resonant caustic degeneracy broken}",
      journal = {\mnras},
         year = 2023,
        month = jul,
       volume = {522},
       number = {4},
        pages = {6055-6069},
          doi = {10.1093/mnras/stad1398},
archivePrefix = {arXiv},
       eprint = {2301.06779},
 primaryClass = {astro-ph.EP},
       adsurl = {https://ui.adsabs.harvard.edu/abs/2023MNRAS.522.6055Z}
}

@ARTICLE{OB050071,
   author = {{Udalski}, A. and {Jaroszy{\'n}ski}, M. and {Paczy{\'n}ski}, B. and 
	{Kubiak}, M. and {Szyma{\'n}ski}, M.~K. and {Soszy{\'n}ski}, I. and 
	{Pietrzy{\'n}ski}, G. and {Ulaczyk}, K. and {Szewczyk}, O. and 
	{Wyrzykowski}, {\L}. and {OGLE Collaboration} and {Christie}, G.~W. and 
	{DePoy}, D.~L. and {Dong}, S. and {Gal-Yam}, A. and {Gaudi}, B.~S. and 
	{Gould}, A. and {Han}, C. and {L{\'e}pine}, S. and {McCormick}, J. and 
	{Park}, B.-G. and {Pogge}, R.~W. and {{$\mu$}FUN Collaboration} and 
	{Bennett}, D.~P. and {Bond}, I.~A. and {Muraki}, Y. and {Tristram}, P.~J. and 
	{Yock}, P.~C.~M. and {MOA Collaboration} and {Beaulieu}, J.-P. and 
	{Bramich}, D.~M. and {Dieters}, S.~W. and {Greenhill}, J. and 
	{Hill}, K. and {Horne}, K. and {Kubas}, D. and {PLANET/ROBONET  Collaboration}
	},
    title = "{A Jovian-Mass Planet in Microlensing Event OGLE-2005-BLG-071}",
  journal = {\apjl},
   eprint = {astro-ph/0505451},
     year = 2005,
    month = aug,
   volume = 628,
    pages = {L109-L112},
      doi = {10.1086/432795},
   adsurl = {http://adsabs.harvard.edu/abs/2005ApJ...628L.109U}
}

@ARTICLE{Bensby2013,
   author = {{Bensby}, T. and {Yee}, J.~C. and {Feltzing}, S. and {Johnson}, J.~A. and 
	{Gould}, A. and {Cohen}, J.~G. and {Asplund}, M. and {Mel{\'e}ndez}, J. and 
	{Lucatello}, S. and {Han}, C. and {Thompson}, I. and {Gal-Yam}, A. and 
	{Udalski}, A. and {Bennett}, D.~P. and {Bond}, I.~A. and {Kohei}, W. and 
	{Sumi}, T. and {Suzuki}, D. and {Suzuki}, K. and {Takino}, S. and 
	{Tristram}, P. and {Yamai}, N. and {Yonehara}, A.},
    title = "{Chemical evolution of the Galactic bulge as traced by microlensed dwarf and subgiant stars. V. Evidence for a wide age distribution and a complex MDF}",
  journal = {\aap},
archivePrefix = "arXiv",
   eprint = {1211.6848},
     year = 2013,
    month = jan,
   volume = 549,
      eid = {A147},
    pages = {A147},
      doi = {10.1051/0004-6361/201220678},
   adsurl = {http://adsabs.harvard.edu/abs/2013A%26A...549A.147B}
}

@ARTICLE{Gould1992,
   author = {{Gould}, A.},
    title = "{Extending the MACHO search to about 10 exp 6 solar masses}",
  journal = {\apj},
     year = 1992,
    month = jun,
   volume = 392,
    pages = {442-451},
      doi = {10.1086/171443},
   adsurl = {http://adsabs.harvard.edu/abs/1992ApJ...392..442G}
}

@ARTICLE{Griest1998,
   author = {{Griest}, K. and {Safizadeh}, N.},
    title = "{The Use of High-Magnification Microlensing Events in Discovering Extrasolar Planets}",
  journal = {\apj},
   eprint = {astro-ph/9710342},
     year = 1998,
    month = jun,
   volume = 500,
    pages = {37-50},
      doi = {10.1086/305729},
   adsurl = {http://adsabs.harvard.edu/abs/1998ApJ...500...37G}
}

@ARTICLE{mufun,
   author = {{Gould}, A. and {Dong}, S. and {Gaudi}, B.~S. and {Udalski}, A. and 
	{Bond}, I.~A. and {Greenhill}, J. and {Street}, R.~A. and {Dominik}, M. and 
	{Sumi}, T. and {Szyma{\'n}ski}, M.~K. and {Han}, C. and {Allen}, W. and 
	{Bolt}, G. and {Bos}, M. and {Christie}, G.~W. and {DePoy}, D.~L. and 
	{Drummond}, J. and {Eastman}, J.~D. and {Gal-Yam}, A. and {Higgins}, D. and 
	{Janczak}, J. and {Kaspi}, S. and {Koz{\l}owski}, S. and {Lee}, C.-U. and 
	{Mallia}, F. and {Maury}, A. and {Maoz}, D. and {McCormick}, J. and 
	{Monard}, L.~A.~G. and {Moorhouse}, D. and {Morgan}, N. and 
	{Natusch}, T. and {Ofek}, E.~O. and {Park}, B.-G. and {Pogge}, R.~W. and 
	{Polishook}, D. and {Santallo}, R. and {Shporer}, A. and {Spector}, O. and 
	{Thornley}, G. and {Yee}, J.~C. and {{$\mu$}FUN Collaboration} and 
	{Kubiak}, M. and {Pietrzy{\'n}ski}, G. and {Soszy{\'n}ski}, I. and 
	{Szewczyk}, O. and {Wyrzykowski}, {\L}. and {Ulaczyk}, K. and 
	{Poleski}, R. and {OGLE Collaboration} and {Abe}, F. and {Bennett}, D.~P. and 
	{Botzler}, C.~S. and {Douchin}, D. and {Freeman}, M. and {Fukui}, A. and 
	{Furusawa}, K. and {Hearnshaw}, J.~B. and {Hosaka}, S. and {Itow}, Y. and 
	{Kamiya}, K. and {Kilmartin}, P.~M. and {Korpela}, A. and {Lin}, W. and 
	{Ling}, C.~H. and {Makita}, S. and {Masuda}, K. and {Matsubara}, Y. and 
	{Miyake}, N. and {Muraki}, Y. and {Nagaya}, M. and {Nishimoto}, K. and 
	{Ohnishi}, K. and {Okumura}, T. and {Perrott}, Y.~C. and {Philpott}, L. and 
	{Rattenbury}, N. and {Saito}, T. and {Sako}, T. and {Sullivan}, D.~J. and 
	{Sweatman}, W.~L. and {Tristram}, P.~J. and {von Seggern}, E. and 
	{Yock}, P.~C.~M. and {MOA Collaboration} and {Albrow}, M. and 
	{Batista}, V. and {Beaulieu}, J.~P. and {Brillant}, S. and {Caldwell}, J. and 
	{Calitz}, J.~J. and {Cassan}, A. and {Cole}, A. and {Cook}, K. and 
	{Coutures}, C. and {Dieters}, S. and {Dominis Prester}, D. and 
	{Donatowicz}, J. and {Fouqu{\'e}}, P. and {Hill}, K. and {Hoffman}, M. and 
	{Jablonski}, F. and {Kane}, S.~R. and {Kains}, N. and {Kubas}, D. and 
	{Marquette}, J.-B. and {Martin}, R. and {Martioli}, E. and {Meintjes}, P. and 
	{Menzies}, J. and {Pedretti}, E. and {Pollard}, K. and {Sahu}, K.~C. and 
	{Vinter}, C. and {Wambsganss}, J. and {Watson}, R. and {Williams}, A. and 
	{Zub}, M. and {PLANET Collaboration} and {Allan}, A. and {Bode}, M.~F. and 
	{Bramich}, D.~M. and {Burgdorf}, M.~J. and {Clay}, N. and {Fraser}, S. and 
	{Hawkins}, E. and {Horne}, K. and {Kerins}, E. and {Lister}, T.~A. and 
	{Mottram}, C. and {Saunders}, E.~S. and {Snodgrass}, C. and 
	{Steele}, I.~A. and {Tsapras}, Y. and {RoboNet Collaboration} and 
	{J{\o}rgensen}, U.~G. and {Anguita}, T. and {Bozza}, V. and 
	{Calchi Novati}, S. and {Harps{\o}e}, K. and {Hinse}, T.~C. and 
	{Hundertmark}, M. and {Kj{\ae}rgaard}, P. and {Liebig}, C. and 
	{Mancini}, L. and {Masi}, G. and {Mathiasen}, M. and {Rahvar}, S. and 
	{Ricci}, D. and {Scarpetta}, G. and {Southworth}, J. and {Surdej}, J. and 
	{Th{\"o}ne}, C.~C. and {MiNDSTEp Consortium}},
    title = "{Frequency of Solar-like Systems and of Ice and Gas Giants Beyond the Snow Line from High-magnification Microlensing Events in 2005-2008}",
  journal = {\apj},
archivePrefix = "arXiv",
   eprint = {1001.0572},
 primaryClass = "astro-ph.EP",
     year = 2010,
    month = sep,
   volume = 720,
    pages = {1073-1089},
      doi = {10.1088/0004-637X/720/2/1073},
   adsurl = {http://adsabs.harvard.edu/abs/2010ApJ...720.1073G}
}

@ARTICLE{An2002,
   author = {{An}, J.~H. and {Albrow}, M.~D. and {Beaulieu}, J.-P. and {Caldwell}, J.~A.~R. and 
	{DePoy}, D.~L. and {Dominik}, M. and {Gaudi}, B.~S. and {Gould}, A. and 
	{Greenhill}, J. and {Hill}, K. and {Kane}, S. and {Martin}, R. and 
	{Menzies}, J. and {Pogge}, R.~W. and {Pollard}, K.~R. and {Sackett}, P.~D. and 
	{Sahu}, K.~C. and {Vermaak}, P. and {Watson}, R. and {Williams}, A.
	},
    title = "{First Microlens Mass Measurement: PLANET Photometry of EROS BLG-2000-5}",
  journal = {\apj},
   eprint = {astro-ph/0110095},
     year = 2002,
    month = jun,
   volume = 572,
    pages = {521-539},
      doi = {10.1086/340191},
   adsurl = {http://adsabs.harvard.edu/abs/2002ApJ...572..521A}
}

@ARTICLE{Bessell1988,
   author = {{Bessell}, M.~S. and {Brett}, J.~M.},
    title = "{JHKLM photometry - Standard systems, passbands, and intrinsic colors}",
  journal = {\pasp},
     year = 1988,
    month = sep,
   volume = 100,
    pages = {1134-1151},
      doi = {10.1086/132281},
   adsurl = {http://adsabs.harvard.edu/abs/1988PASP..100.1134B}
}

@ARTICLE{Kroupa2001,
   author = {{Kroupa}, P.},
    title = "{On the variation of the initial mass function}",
  journal = {\mnras},
   eprint = {astro-ph/0009005},
     year = 2001,
    month = apr,
   volume = 322,
    pages = {231-246},
      doi = {10.1046/j.1365-8711.2001.04022.x},
   adsurl = {http://adsabs.harvard.edu/abs/2001MNRAS.322..231K}
}

@ARTICLE{Yang_TLC,
       author = {{Yang}, Hongjing and {Yee}, Jennifer C. and {Hwang}, Kyu-Ha and {Qian}, Qiyue and {Bond}, Ian A. and {Gould}, Andrew and {Hu}, Zhecheng and {Zhang}, Jiyuan and {Mao}, Shude and {Zhu}, Wei and {Albrow}, Michael D. and {Chung}, Sun-Ju and {Kim}, Seung-Lee and {Park}, Byeong-Gon and {Han}, Cheongho and {Jung}, Youn Kil and {Ryu}, Yoon-Hyun and {Shin}, In-Gu and {Shvartzvald}, Yossi and {Cha}, Sang-Mok and {Kim}, Dong-Jin and {Kim}, Hyoun-Woo and {Lee}, Chung-Uk and {Lee}, Dong-Joo and {Lee}, Yongseok and {Pogge}, Richard W. and {Zang}, Weicheng and {Abe}, Fumio and {Barry}, Richard and {Bennett}, David P. and {Bhattacharya}, Aparna and {Donachie}, Martin and {Fujii}, Hirosane and {Fukui}, Akihiko and {Hirao}, Yuki and {Itow}, Yoshitaka and {Kirikawa}, Rintaro and {Kondo}, Iona and {Koshimoto}, Naoki and {Silva}, Stela Ishitani and {Li}, Man Cheung Alex and {Matsubara}, Yutaka and {Muraki}, Yasushi and {Suzuki}, Daisuke and {Tristram}, Paul J. and {Yonehara}, Atsunori and {Ranc}, Cl{\'e}ment and {Miyazaki}, Shota and {Olmschenk}, Greg and {Rattenbury}, Nicholas J. and {Satoh}, Yuki and {Shoji}, Hikaru and {Sumi}, Takahiro and {Tanaka}, Yuzuru and {Yamawaki}, Tsubasa},
        title = "{Systematic reanalysis of KMTNet microlensing events, paper I: Updates of the photometry pipeline and a new planet candidate}",
      journal = {\mnras},
         year = 2024,
        month = feb,
       volume = {528},
       number = {1},
        pages = {11-27},
          doi = {10.1093/mnras/stad3672},
archivePrefix = {arXiv},
       eprint = {2311.04876},
 primaryClass = {astro-ph.EP},
       adsurl = {https://ui.adsabs.harvard.edu/abs/2024MNRAS.528...11Y}
}

@ARTICLE{Yang_TLC2,
       author = {{Yang}, Hongjing and {Yee}, Jennifer C. and {Zhang}, Jiyuan and {Lee}, Chung-Uk and {Kim}, Dong-Jin and {Bond}, Ian A. and {Udalski}, Andrzej and {Hwang}, Kyu-Ha and {Zang}, Weicheng and {Qian}, Qiyue and {Gould}, Andrew and {Mao}, Shude and {Albrow}, Michael D. and {Chung}, Sun-Ju and {Han}, Cheongho and {Jung}, Youn Kil and {Ryu}, Yoon-Hyun and {Shin}, In-Gu and {Shvartzvald}, Yossi and {Cha}, Sang-Mok and {Kim}, Hyoun-Woo and {Kim}, Seung-Lee and {Lee}, Dong-Joo and {Lee}, Yongseok and {Park}, Byeong-Gon and {Pogge}, Richard W. and {KMTNet Collaboration} and {Abe}, Fumio and {Bando}, Ken and {Bennett}, David P. and {Bhattacharya}, Aparna and {Fukui}, Akihiko and {Hamada}, Ryusei and {Hamada}, Shunya and {Hamasaki}, Naoto and {Hirao}, Yuki and {Ishitani Silva}, Stela and {Itow}, Yoshitaka and {Koshimoto}, Naoki and {Matsubara}, Yutaka and {Miyazaki}, Shota and {Muraki}, Yasushi and {Nagai}, Tutumi and {Nunota}, Kansuke and {Olmschenk}, Greg and {Ranc}, Cl{\'e}ment and {Rattenbury}, Nicholas J. and {Satoh}, Yuki and {Sumi}, Takahiro and {Suzuki}, Daisuke and {Terry}, Sean K. and {Tristram}, Paul. J. and {Vandorou}, Aikaterini and {Yama}, Hibiki and {MOA Collaboration} and {Mr{\'o}z}, Przemek and {Skowron}, Jan and {Poleski}, Radoslaw and {Szyma{\'n}ski}, Micha{\l} K. and {Soszy{\'n}ski}, Igor and {Pietrukowicz}, Pawe{\l} and {Koz{\l}owski}, Szymon and {Ulaczyk}, Krzysztof and {Rybicki}, Krzysztof A. and {Iwanek}, Patryk and {Wrona}, Marcin and {OGLE Collaboration}},
        title = "{Systematic Reanalysis of KMTNet Microlensing Events. II. Two New Planets in Giant-source Events}",
      journal = {\aj},
         year = 2025,
        month = jun,
       volume = {169},
       number = {6},
          eid = {295},
        pages = {295},
          doi = {10.3847/1538-3881/adc73e},
archivePrefix = {arXiv},
       eprint = {2503.19471},
 primaryClass = {astro-ph.EP},
       adsurl = {https://ui.adsabs.harvard.edu/abs/2025AJ....169..295Y}
}

@ARTICLE{Suzuki2016,
       author = {{Suzuki}, D. and {Bennett}, D.~P. and {Sumi}, T. and {Bond}, I.~A. and
         {Rogers}, L.~A. and {Abe}, F. and {Asakura}, Y. and {Bhattacharya}, A. and
         {Donachie}, M. and {Freeman}, M. and {Fukui}, A. and {Hirao}, Y. and
         {Itow}, Y. and {Koshimoto}, N. and {Li}, M.~C.~A. and {Ling}, C.~H. and
         {Masuda}, K. and {Matsubara}, Y. and {Muraki}, Y. and {Nagakane}, M. and
         {Onishi}, K. and {Oyokawa}, H. and {Rattenbury}, N. and {Saito}, To. and
         {Sharan}, A. and {Shibai}, H. and {Sullivan}, D.~J. and
         {Tristram}, P.~J. and {Yonehara}, A. and {MOA Collaboration}},
        title = "{The Exoplanet Mass-ratio Function from the MOA-II Survey: Discovery of a Break and Likely Peak at a Neptune Mass}",
      journal = {\apj},
         year = "2016",
        month = "Dec",
       volume = {833},
          eid = {145},
        pages = {145},
          doi = {10.3847/1538-4357/833/2/145},
archivePrefix = {arXiv},
       eprint = {1612.03939},
 primaryClass = {astro-ph.EP},
       adsurl = {https://ui.adsabs.harvard.edu/\#abs/2016ApJ...833..145S}
}

@ARTICLE{OB09020,
   author = {{Skowron}, J. and {Udalski}, A. and {Gould}, A. and {Dong}, S. and 
	{Monard}, L.~A.~G. and {Han}, C. and {Nelson}, C.~R. and {McCormick}, J. and 
	{Moorhouse}, D. and {Thornley}, G. and {Maury}, A. and {Bramich}, D.~M. and 
	{Greenhill}, J. and {Koz{\l}owski}, S. and {Bond}, I. and {Poleski}, R. and 
	{Wyrzykowski}, {\L}. and {Ulaczyk}, K. and {Kubiak}, M. and 
	{Szyma{\'n}ski}, M.~K. and {Pietrzy{\'n}ski}, G. and {Soszy{\'n}ski}, I. and 
	{OGLE Collaboration} and {Gaudi}, B.~S. and {Yee}, J.~C. and 
	{Hung}, L.-W. and {Pogge}, R.~W. and {DePoy}, D.~L. and {Lee}, C.-U. and 
	{Park}, B.-G. and {Allen}, W. and {Mallia}, F. and {Drummond}, J. and 
	{Bolt}, G. and {{$\mu$}FUN Collaboration} and {Allan}, A. and 
	{Browne}, P. and {Clay}, N. and {Dominik}, M. and {Fraser}, S. and 
	{Horne}, K. and {Kains}, N. and {Mottram}, C. and {Snodgrass}, C. and 
	{Steele}, I. and {Street}, R.~A. and {Tsapras}, Y. and {RoboNet Collaboration} and 
	{Abe}, F. and {Bennett}, D.~P. and {Botzler}, C.~S. and {Douchin}, D. and 
	{Freeman}, M. and {Fukui}, A. and {Furusawa}, K. and {Hayashi}, F. and 
	{Hearnshaw}, J.~B. and {Hosaka}, S. and {Itow}, Y. and {Kamiya}, K. and 
	{Kilmartin}, P.~M. and {Korpela}, A. and {Lin}, W. and {Ling}, C.~H. and 
	{Makita}, S. and {Masuda}, K. and {Matsubara}, Y. and {Muraki}, Y. and 
	{Nagayama}, T. and {Miyake}, N. and {Nishimoto}, K. and {Ohnishi}, K. and 
	{Perrott}, Y.~C. and {Rattenbury}, N. and {Saito}, T. and {Skuljan}, L. and 
	{Sullivan}, D.~J. and {Sumi}, T. and {Suzuki}, D. and {Sweatman}, W.~L. and 
	{Tristram}, P.~J. and {Wada}, K. and {Yock}, P.~C.~M. and {MOA Collaboration} and 
	{Beaulieu}, J.-P. and {Fouqu{\'e}}, P. and {Albrow}, M.~D. and 
	{Batista}, V. and {Brillant}, S. and {Caldwell}, J.~A.~R. and 
	{Cassan}, A. and {Cole}, A. and {Cook}, K.~H. and {Coutures}, C. and 
	{Dieters}, S. and {Dominis Prester}, D. and {Donatowicz}, J. and 
	{Kane}, S.~R. and {Kubas}, D. and {Marquette}, J.-B. and {Martin}, R. and 
	{Menzies}, J. and {Sahu}, K.~C. and {Wambsganss}, J. and {Williams}, A. and 
	{Zub}, M. and {PLANET Collaboration}},
    title = "{Binary Microlensing Event OGLE-2009-BLG-020 Gives Verifiable Mass, Distance, and Orbit Predictions}",
  journal = {\apj},
archivePrefix = "arXiv",
   eprint = {1101.3312},
 primaryClass = "astro-ph.SR",
     year = 2011,
    month = sep,
   volume = 738,
      eid = {87},
    pages = {87},
      doi = {10.1088/0004-637X/738/1/87},
   adsurl = {http://adsabs.harvard.edu/abs/2011ApJ...738...87S}
}

@ARTICLE{HSTCMD,
   author = {{Holtzman}, J.~A. and {Watson}, A.~M. and {Baum}, W.~A. and 
	{Grillmair}, C.~J. and {Groth}, E.~J. and {Light}, R.~M. and 
	{Lynds}, R. and {O'Neil}, Jr., E.~J.},
    title = "{The Luminosity Function and Initial Mass Function in the Galactic Bulge}",
  journal = {\aj},
   eprint = {astro-ph/9801321},
     year = 1998,
    month = may,
   volume = 115,
    pages = {1946-1957},
      doi = {10.1086/300336},
   adsurl = {http://adsabs.harvard.edu/abs/1998AJ....115.1946H}
}

@ARTICLE{MB07192,
   author = {{Bennett}, D.~P. and {Bond}, I.~A. and {Udalski}, A. and {Sumi}, T. and 
	{Abe}, F. and {Fukui}, A. and {Furusawa}, K. and {Hearnshaw}, J.~B. and 
	{Holderness}, S. and {Itow}, Y. and {Kamiya}, K. and {Korpela}, A.~V. and 
	{Kilmartin}, P.~M. and {Lin}, W. and {Ling}, C.~H. and {Masuda}, K. and 
	{Matsubara}, Y. and {Miyake}, N. and {Muraki}, Y. and {Nagaya}, M. and 
	{Okumura}, T. and {Ohnishi}, K. and {Perrott}, Y.~C. and {Rattenbury}, N.~J. and 
	{Sako}, T. and {Saito}, T. and {Sato}, S. and {Skuljan}, L. and 
	{Sullivan}, D.~J. and {Sweatman}, W.~L. and {Tristram}, P.~J. and 
	{Yock}, P.~C.~M. and {Kubiak}, M. and {Szyma{\'n}ski}, M.~K. and 
	{Pietrzy{\'n}ski}, G. and {Soszy{\'n}ski}, I. and {Szewczyk}, O. and 
	{Wyrzykowski}, {\L}. and {Ulaczyk}, K. and {Batista}, V. and 
	{Beaulieu}, J.~P. and {Brillant}, S. and {Cassan}, A. and {Fouqu{\'e}}, P. and 
	{Kervella}, P. and {Kubas}, D. and {Marquette}, J.~B.},
    title = "{A Low-Mass Planet with a Possible Sub-Stellar-Mass Host in Microlensing Event MOA-2007-BLG-192}",
  journal = {\apj},
archivePrefix = "arXiv",
   eprint = {0806.0025},
     year = 2008,
    month = sep,
   volume = 684,
    pages = {663-683},
      doi = {10.1086/589940},
   adsurl = {http://adsabs.harvard.edu/abs/2008ApJ...684..663B}
}

@ARTICLE{MatthewWFIRSTI,
       author = {{Penny}, Matthew T. and {Gaudi}, B. Scott and {Kerins}, Eamonn and
         {Rattenbury}, Nicholas J. and {Mao}, Shude and {Robin}, Annie C. and
         {Calchi Novati}, Sebastiano},
        title = "{Predictions of the WFIRST Microlensing Survey. I. Bound Planet Detection Rates}",
      journal = {\apjs},
         year = "2019",
        month = "Mar",
       volume = {241},
       number = {1},
          eid = {3},
        pages = {3},
          doi = {10.3847/1538-4365/aafb69},
archivePrefix = {arXiv},
       eprint = {1808.02490},
 primaryClass = {astro-ph.EP},
       adsurl = {https://ui.adsabs.harvard.edu/abs/2019ApJS..241....3P}
}

@ARTICLE{KMTAF,
       author = {{Kim}, Hyoun-Woo and {Hwang}, Kyu-Ha and {Shvartzvald}, Yossi and
         {Yee}, Jennifer C. and {Albrow}, Michael D. and {Cha}, Sang-Mok and
         {Chung}, Sun-Ju and {Gould}, Andrew and {Han}, Cheongho and
         {Jung}, Youn Kil and {Kim}, Dong-Jin and {Kim}, Seung-Lee and
         {Lee}, Chung-Uk and {Lee}, Dong-Joo and {Lee}, Yongseok and
         {Park}, Byeong-Gon and {Pogge}, Richard W. and {Ryu}, Yoon-Hyun and
         {Shin}, In-Gu and {Zang}, Weicheng},
        title = "{The Korea Microlensing Telescope Network (KMTNet) Alert Algorithm and Alert System}",
      journal = {arXiv e-prints},
         year = 2018,
        month = jun,
          eid = {arXiv:1806.07545},
        pages = {arXiv:1806.07545},
archivePrefix = {arXiv},
       eprint = {1806.07545},
 primaryClass = {astro-ph.IM},
       adsurl = {https://ui.adsabs.harvard.edu/abs/2018arXiv180607545K}
}

@ARTICLE{MB11293,
       author = {{Yee}, J.~C. and {Shvartzvald}, Y. and {Gal-Yam}, A. and {Bond}, I.~A. and
         {Udalski}, A. and {Koz{\l}owski}, S. and {Han}, C. and {Gould}, A. and
         {Skowron}, J. and {Suzuki}, D. and {Abe}, F. and {Bennett}, D.~P. and
         {Botzler}, C.~S. and {Chote}, P. and {Freeman}, M. and {Fukui}, A. and
         {Furusawa}, K. and {Itow}, Y. and {Kobara}, S. and {Ling}, C.~H. and
         {Masuda}, K. and {Matsubara}, Y. and {Miyake}, N. and {Muraki}, Y. and
         {Ohmori}, K. and {Ohnishi}, K. and {Rattenbury}, N.~J. and
         {Saito}, To. and {Sullivan}, D.~J. and {Sumi}, T. and {Suzuki}, K. and
         {Sweatman}, W.~L. and {Takino}, S. and {Tristram}, P.~J. and
         {Wada}, K. and {MOA Collaboration} and {Szyma{\'n}ski}, M.~K. and
         {Kubiak}, M. and {Pietrzy{\'n}ski}, G. and {Soszy{\'n}ski}, I. and
         {Poleski}, R. and {Ulaczyk}, K. and {Wyrzykowski}, {\L}. and
         {Pietrukowicz}, P. and {OGLE Collaboration} and {Allen}, W. and
         {Almeida}, L.~A. and {Batista}, V. and {Bos}, M. and {Christie}, G. and
         {DePoy}, D.~L. and {Dong}, Subo and {Drummond}, J. and {Finkelman}, I. and
         {Gaudi}, B.~S. and {Gorbikov}, E. and {Henderson}, C. and
         {Higgins}, D. and {Jablonski}, F. and {Kaspi}, S. and {Manulis}, I. and
         {Maoz}, D. and {McCormick}, J. and {McGregor}, D. and
         {Monard}, L.~A.~G. and {Moorhouse}, D. and {Mu{\~n}oz}, J.~A. and
         {Natusch}, T. and {Ngan}, H. and {Ofek}, E. and {Pogge}, R.~W. and
         {Santallo}, R. and {Tan}, T. -G. and {Thornley}, G. and {Shin}, I. -G. and
         {Choi}, J. -Y. and {Park}, S. -Y. and {Lee}, C. -U. and {Koo}, J. -R. and
         {{\ensuremath{\mu}}FUN Collaboration}},
        title = "{MOA-2011-BLG-293Lb: A Test of Pure Survey Microlensing Planet Detections}",
      journal = {\apj},
         year = 2012,
        month = aug,
       volume = {755},
       number = {2},
          eid = {102},
        pages = {102},
          doi = {10.1088/0004-637X/755/2/102},
archivePrefix = {arXiv},
       eprint = {1201.1002},
 primaryClass = {astro-ph.EP},
       adsurl = {https://ui.adsabs.harvard.edu/abs/2012ApJ...755..102Y}
}

@ARTICLE{Tomaney1996,
       author = {{Tomaney}, Austin B. and {Crotts}, Arlin P.~S.},
        title = "{Expanding the Realm of Microlensing Surveys with Difference Image Photometry}",
      journal = {\aj},
         year = "1996",
        month = "Dec",
       volume = {112},
        pages = {2872},
          doi = {10.1086/118228},
archivePrefix = {arXiv},
       eprint = {astro-ph/9610066},
 primaryClass = {astro-ph},
       adsurl = {https://ui.adsabs.harvard.edu/abs/1996AJ....112.2872T}
}

@ARTICLE{Wise,
       author = {{Shvartzvald}, Y. and {Maoz}, D. and {Udalski}, A. and {Sumi}, T. and
         {Friedmann}, M. and {Kaspi}, S. and {Poleski}, R. and
         {Szyma{\'n}ski}, M.~K. and {Skowron}, J. and {Koz{\l}owski}, S. and
         {Wyrzykowski}, {\L}. and {Mr{\'o}z}, P. and {Pietrukowicz}, P. and
         {Pietrzy{\'n}ski}, G. and {Soszy{\'n}ski}, I. and {Ulaczyk}, K. and
         {Abe}, F. and {Barry}, R.~K. and {Bennett}, D.~P. and
         {Bhattacharya}, A. and {Bond}, I.~A. and {Freeman}, M. and
         {Inayama}, K. and {Itow}, Y. and {Koshimoto}, N. and {Ling}, C.~H. and
         {Masuda}, K. and {Fukui}, A. and {Matsubara}, Y. and {Muraki}, Y. and
         {Ohnishi}, K. and {Rattenbury}, N.~J. and {Saito}, To. and
         {Sullivan}, D.~J. and {Suzuki}, D. and {Tristram}, P.~J. and
         {Wakiyama}, Y. and {Yonehara}, A.},
        title = "{The frequency of snowline-region planets from four years of OGLE-MOA-Wise second-generation microlensing}",
      journal = {\mnras},
         year = 2016,
        month = apr,
       volume = {457},
       number = {4},
        pages = {4089-4113},
          doi = {10.1093/mnras/stw191},
archivePrefix = {arXiv},
       eprint = {1510.04297},
 primaryClass = {astro-ph.EP},
       adsurl = {https://ui.adsabs.harvard.edu/abs/2016MNRAS.457.4089S}
}

@ARTICLE{OB110173,
       author = {{Poleski}, Rados{\l}aw and {Gaudi}, B.~S. and {Udalski}, A. and
         {Szyma{\'n}ski}, M.~K. and {Soszy{\'n}ski}, I. and {Pietrukowicz}, P. and
         {Koz{\l}owski}, S. and {Skowron}, J. and {Wyrzykowski}, {\L}. and
         {Ulaczyk}, K.},
        title = "{An Ice Giant Exoplanet Interpretation of the Anomaly in Microlensing Event OGLE-2011-BLG-0173}",
      journal = {\aj},
         year = 2018,
        month = sep,
       volume = {156},
       number = {3},
          eid = {104},
        pages = {104},
          doi = {10.3847/1538-3881/aad45e},
archivePrefix = {arXiv},
       eprint = {1805.00049},
 primaryClass = {astro-ph.EP},
       adsurl = {https://ui.adsabs.harvard.edu/abs/2018AJ....156..104P}
}

@ARTICLE{Mordasini2009,
       author = {{Mordasini}, C. and {Alibert}, Y. and {Benz}, W.},
        title = "{Extrasolar planet population synthesis. I. Method, formation tracks, and mass-distance distribution}",
      journal = {\aap},
         year = 2009,
        month = jul,
       volume = {501},
       number = {3},
        pages = {1139-1160},
          doi = {10.1051/0004-6361/200810301},
archivePrefix = {arXiv},
       eprint = {0904.2524},
 primaryClass = {astro-ph.EP},
       adsurl = {https://ui.adsabs.harvard.edu/abs/2009A&A...501.1139M}
}

@ARTICLE{Bozza2018,
       author = {{Bozza}, Valerio and {Bachelet}, Etienne and {Bartoli{\'c}}, Fran and
         {Heintz}, Tyler M. and {Hoag}, Ava R. and {Hundertmark}, Markus},
        title = "{VBBINARYLENSING: a public package for microlensing light-curve computation}",
      journal = {\mnras},
         year = 2018,
        month = oct,
       volume = {479},
       number = {4},
        pages = {5157-5167},
          doi = {10.1093/mnras/sty1791},
archivePrefix = {arXiv},
       eprint = {1805.05653},
 primaryClass = {astro-ph.IM},
       adsurl = {https://ui.adsabs.harvard.edu/abs/2018MNRAS.479.5157B}
}

\end{CJK*}
\end{document}